\documentclass[referee]{aa} % for a referee version
\usepackage{bm}
\usepackage{amsmath}
\usepackage{natbib}
\usepackage[a4wide]{}
\usepackage[english]{babel}
\usepackage{graphicx}
\usepackage{textcomp}
\usepackage{subcaption}
\usepackage{lscape}
\usepackage{color}
\usepackage{verbatim} 
\usepackage{hyperref}
\usepackage{widetext}

\def\bk{{\bm k}}

\def\bn{{\bm n}}
\def\bN{{\bm N}}

\def\bx{{\bm x}}

\def\rab{\mbox{ $R_{AB}$}}

\def\nn{\bN^{(1)}.\bN^{(2)}}

\def\bk{{\bm k}}

\def\bn{{\bm n}}

\def\bx{{\bm x}}

\def\bN{{\bm N}}

\def\rab{R_{AB}}

\def\lb{\label}
\def\be{\begin{equation}}
\def\ee{\end{equation}}
\def\bea{\begin{eqnarray}}
\def\eea{\end{eqnarray}}

\newcommand{\abs}[1]{\left | #1 \right |}

\def\lb{\label}

\def\be{\begin{equation}}
\def\ee{\end{equation}}
\def\bea{\begin{eqnarray}}
\def\eea{\end{eqnarray}}

\def\bx{{\bm x}}

\newcommand{\norm}[1]{|{#1}|}

\def\lb{\label}

\def\brpa{\bm{R_{PA}}}

\def\rpa{R_{PA}}
\def\rpb{R_{PB}}

\def\bnab{\bm N_{AB}}

\def\bnpa{\bm N_{PA}}
\def\bnpb{\bm N_{PB}}

\def\nn{\nonumber}

\def\be{\begin{equation}}
\def\ee{\end{equation}}
\def\bea{\begin{eqnarray}}
\def\eea{\end{eqnarray}}

\def\nabi{N_{AB}^i}

\def\bsube{\begin{subequations}}
\def\esube{\end{subequations}}

\def\aap{{Astronomy \& Astrophysics}}

\begin{document}

\title{Application of Time Transfer Functions to Gaia's global astrometry}
\titlerunning{Application of the TTF to Gaia's global astrometry}

\subtitle{%Numerical validation and sphere reconstruction from Gaia simulated data
Validation on DPAC simulated Gaia-like observations}

\author{Stefano~Bertone\inst{1,2} \thanks{\emph{Present address:}
    Astronomical Institute, University of Bern, Sidlerstrasse 5, CH-3011, Bern, Switzerland} \and Alberto~Vecchiato\inst{1} \and Beatrice~Bucciarelli\inst{1,3} \and Mariateresa~Crosta\inst{1} \and Mario~G.~Lattanzi\inst{1} \and Luca~Bianchi\inst{4} \and Marie-Christine~Angonin\inst{2} \and Christophe~Le~Poncin-Lafitte\inst{2}
    }
\authorrunning{S. Bertone et al}

\institute{INAF, Astrophysical Observatory of Torino, Via Osservatorio 20, 10025 Pino Torinese (Torino), Italy
\and
SYRTE, Observatoire de Paris, PSL Research University, CNRS, Sorbonne Universit\'es, UPMC Univ. Paris 06, LNE, 61 avenue de
l'Observatoire, 75014 Paris, France
\and
Shanghai Astronomical Observatory, Chinese Academy of Sciences, 80 Nandan Road, Shanghai 200030, China
\and
EURIX SRL, via Giulio Carcano 26, 10153 Torino, Italy}

   \date{\today}

% \abstract{}{}{}{}{} 
% 5 {} token are mandatory
 
  \abstract
   {%The Gaia space satellite is currently performing absolute astrometry, aiming at the definition of a global astrometric reference frame at visual wavelengths. This requires an accurate modeling of relativistic effects on light propagation, as well as procedures for the internal validation of the results.
   A key objective of the ESA Gaia satellite is the realization of a quasi-inertial 
reference frame at visual wavelengths by means of global astrometric techniques. This requires an 
accurate mathematical and numerical modeling of relativistic light propagation, as well as 
double-blind-like procedures for the internal validation of the results, before they are released 
to the scientific community at large.
   }
   {%We apply the Time Transfer Functions (TTF) formalism to the solution of the global astrometric problem to prove its performances and to provide an additional external validation option for Gaia solutions and final catalog.
   Specialize the Time Transfer Functions (TTF) formalism to the case of the Gaia observer and 
prove its applicability to the task of Global Sphere Reconstruction (GSR), in anticipation of its
inclusion in the GSR system, already featuring the suite of RAMOD models, as an additional 
semi-external validation of the forthcoming Gaia baseline astrometric solutions.
   }
   {%We extend the current Global Sphere Reconstruction (GSR) framework by introducing the TTF as additional option to compute relativistic light propagation in the Solar System. We compare our results for the computed direction of the observed stars to the results of the Gaia RElativistic Model (GREM). Finally, we setup the solution of a stellar sphere using the TTF along with the functionalities provided within GSR and a set of simulated observations with synthetic measurement errors.
   Extend the current GSR framework and software infrastructure (GSR2) to include TTF
relativistic observation equations compatible with Gaia's operations. Use simulated data generated 
by the Gaia Data Reduction and Analysis Consortium (DPAC) to obtain different least-squares 
estimations of the full (5-parameter) stellar spheres and gauge results. These are compared to
analogous solutions obtained with the current RAMOD model in GSR2 (RAMOD@GSR2) and to the catalog 
generated with the Gaia RElativistic Model (GREM), the model baselined for Gaia and used to 
generate the DPAC synthetic data.
}
   {%Our approach and implementation are shown to be equivalent to the official one used by the Gaia Consortium at Gaia nominal accuracy. The solution of a sphere of about $132,000$ stars based on simulated data and uniformly distributed on the sky shows the potential of this approach for the application to real data.
   Linearized least-squares TTF solutions are based on spheres of about 132,000 primary stars uniformly distributed on the sky and simulated observations spanning the entire 5-yr range of Gaia's nominal operational lifetime. The statistical properties of the results compare well with those of GREM. Finally, comparisons to RAMOD@GSR2 solutions confirmed the known lower accuracy of that model and allowed us to establish firm limits on the quality of the linearization point outside of which an iteration for non-linearity is required for its proper convergence. This has proved invaluable as RAMOD@GSR2 is prepared to go into operations on real satellite data.
}
   {}


%\begin{abstract} 
%\red{\begin{description} 
%	\item[Background] A lot of fundamental tests of gravitational theories rest on highly precise measurements of the travel time and/or the frequency shift of electromagnetic signals propagating through the gravitational field of the Solar System. Moreover, future space astrometry missions, such as Gaia, will soon provide large astrometric catalogues with typical accuracy of several microarcseconds for which a precise relativistic modeling of light propagation is needed. 
%	\item[Purpose] It has been recently demonstrated that this task is not at all mandatory and can be replaced by another approach, initially based on the time transfer functions. 
% 	\item[Results] In this article we develop this formalism for a metric generated by moving gravitational sources and compare our results to the ones given by some geodesic approaches.
%	\item[Conclusions] + motivations for development with a dynamical metric, etc.. 
%\end{description}	
% }
%\end{abstract}

\maketitle

%%%%%%%%%%%%%%%%%%%%%%%%%%%%%%%%%%%%%%%%%%%%%%%%%%%%%%%%%%%%%%%%%%%%%%%%%%
\section{Introduction}
%%%%%%%%%%%%%%%%%%%%%%%%%%%%%%%%%%%%%%%%%%%%%%%%%%%%%%%%%%%%%%%%%%%%%%%%%%
The Gaia space satellite, operational since 2014, performs $4\Pi$ absolute astrometry, aiming at the definition of a global astrometric reference frame at visual wavelengths to unprecedented accuracies. Within this context, it will be very difficult to identify possible errors in the measurements or in the data reduction process by using external comparisons of similar accuracies across the sky as they are simply not available. For this reason, beside the main processing pipeline, the Gaia Data Processing and Analysis Consortium (DPAC) has established an Astrometric Verification Unit (AVU) to verify crucial steps in the baseline data processing chain and report on any significant difference~\citep{2012SPIE.8451E..3CV}. This step is essential in order to guarantee a high quality catalog to the larger scientific community.

Moreover, the processing of astrometric observations at the $\mu$as accuracy required by the Gaia mission demands to take into account relativistic effects on clock synchronization, reference frame transformations as well as on light propagation. Indeed, the behavior of Electromagnetic Waves (EW) in the Solar System is highly sensitive to space-time curvature. 
Most of the available models are based on the solution of null geodesic equations, including those currently used to process Gaia observations: GREM~\citep{2003AJ....125.1580K} for the main processing pipeline of AGIS~\citep{2012A&A...538A..78L} and the RAMOD family~(\cite{2015CQGra..32p5008C,MTRamod4} and references therein) for the AVU as implemented in the Global Sphere Reconstruction (GSR, \cite{2011EAS....45..127A})
%,~\cite{2011EAS....45..127A,2012SPIE.8451E..3CV}
 subsystem hosted at the Italian data processing center in Torino (DPCT,~\cite{2012SPIE.8451E..0EM}).

%Solving the null geodesic equations in the desired context, allows to get the time of flight of an EW between two points, its frequency shift and its deviation angle.  However, solving the geodesic equations involves several steps that turn out to lead to heavy calculations when dealing with metric that describe accurately the dynamical space-time in the solar system \citep{1992AJ....104..897K,2007PhRvD..75f2002K,2011CQGra..28h5010M,2012PhRvD..86d4007D}.
Nevertheless, other approaches exist, such as the one first based on the Synge World Function \citep{2004CQGra..21.4463L} and then improved with the use of the Time Transfer Functions (TTF, see~\cite{2008CQGra..25n5020T}). Similarly to the other methods this approach provides ``ready-to-use'' analytical formulas to describe light propagation in the field of multiple axisimmetric bodies moving with arbitrary velocities in the post-Minkowskian framework~\citep{2014PhRvD..90h4020H} and up to the second and third order for static perturbing bodies~\citep{2013CQGra..30q5008L, 2014PhRvD..89f4045H}.
The equivalence of the TTF approach has been verified analytically with RAMOD in the static case and with GREM 
%and RAMOD has been verified analytically in both the static axisymmetric one-body multipolar first order approximation \citep{2008PhRvD..77d4029L}, in the static one-body monopolar second order approximation \citep{2002PhRvD..66b4045L,2008CQGra..25n5020T,2010CQGra..27g5015K,2010CQGra..27n5013A,2012CQGra..29x5010T} and 
in the case of slowly moving monopoles~\citep{2014CQGra..31a5021B}. However, equivalence on the analytical side does not guarantee equivalence of the numerical results~\citep{2014JPhCS.490a2241V}.

The task of providing the best possible confidence in the data reduction process is precisely the main responsibility of AVU. To this aim, having the possibility of reducing data with different approaches would surely increase the level of reliability from the point of view of the theoretical understanding of the data.
%The same degree of accuracy is either not available yet or would lead to heavy calculations following the direct computation of the null geodesic equations.
%Also, It has been recently demonstrated that solving the geodesic equations is not mandatory and can be replaced by another approach based on the Time Transfer Functions (TTF) \citep{2008CQGra..25n5020T} 
%\footnote{Initially, the method was based on the Synge World Function \citep{2004CQGra..21.4463L} before being improved with TTF \citep{2008CQGra..25n5020T}.} 
%which has proved more efficient, especially when studing light propagation between two point at finite distance. 

In this paper, we present results of the first application of the TTF approach to global astrometry in the framework of the Gaia mission, which is preparatory to its successive inclusion in the AVU/GSR operational infrastructure next to the models of the RAMOD family.
In section~\ref{sect:astroGSR} we briefly recall the astrometric problem of processing Gaia observations and how it is addressed in GSR.
In section~\ref{sect:GSRTTF} we present our new implementation, GSR-TTF, by first describing how the TTF equations have been integrated in the GSR modeling of Gaia observations at the required accuracy. Then, we provide the equations of the coefficients required for the least-square solution of the sphere.
%, consisting of both astrometric and global parameters. 
We take advantage of the modularity of the GSR software infrastructure, which allows for an easy plugin of, $e.g.$, light propagation equations and reference frames transformations.
Finally, we present in section~\ref{sect:GSRTTFres} the comparison of our modeling to the GREM implementation in GaiaTools \citep{2009LL....MTL-019-01,2009LL....PB-015-02} as well as solutions of a sphere of about $132,000$ stars based on DPAC simulated data.
Section~\ref{sect:concl} suggests possible developments and further applications for the results presented in this work and our concluding remarks.

%%%%%%%%%%%%%%%%%%%%%%%%%%%%%%%%%%%
%\section{Notation and conventions}  \label{sect:notconv}
%%%%%%%%%%%%%%%%%%%%%%%%%%%%%%%%%%%
%
%In this paper $c$ is the speed of light in a vacuum and $G$ is the Newtonian gravitational constant. The Lorentzian metric of space-time $V_4$ is denoted by $g$. The signature adopted for $g$ is $(-+++)$. We suppose that space-time is covered by some global quasi-Galilean coordinate system $(x^\mu)=(x^0,\bx )$, where $x^0=ct$, $t$ being a time coordinate, and $\bx=(x^i)$. We assume that the curves of equations $x^i$ = const are timelike, which means that $g_{00}<0$ anywhere. We employ the vector notation $\ba$ in order to denote $(a^1,a^2,a^3)=(a^i)$. 
%Considering two such quantities $\ba$ and $\bb$ we use $\ba \cdot \bb$ to denote $a^i b^i$ (Einstein convention on repeated indices is used). The quantity $\vert \ba \vert$ stands for the ordinary Euclidean norm of $\ba$. 
%For any quantity $f(x^{\lambda})$, $f_{, \alpha}$ denotes the partial derivative of $f$ with respect to $x^{\alpha}$. The indices in parentheses characterize the order of perturbation. They are set up or down, depending on the convenience. 
%If not otherwise specified, when we write $X_{A}$ and  $X_{B}$ it means that we are considering the value of $X$ at coordinate time ($t_A$) and ($t_B$) respectively. 

%%%%%%%%%%%%%%%%%%%%%%%%%%%%%%%%%%
\section{The astrometric problem in GSR} \label{sect:astroGSR}
%%%%%%%%%%%%%%%%%%%%%%%%%%%%%%%%%%
%Before presenting the details of the implementation of our model in GSR, a brief overview of the astrometric problem for Gaia and how it is treated in the software is necessary. 

\subsection{Astrometric coordinates in Gaia} \label{sect:astroCoord}
\begin{figure}[!htbp]
  \begin{center}
    \includegraphics[width=0.5\textwidth]{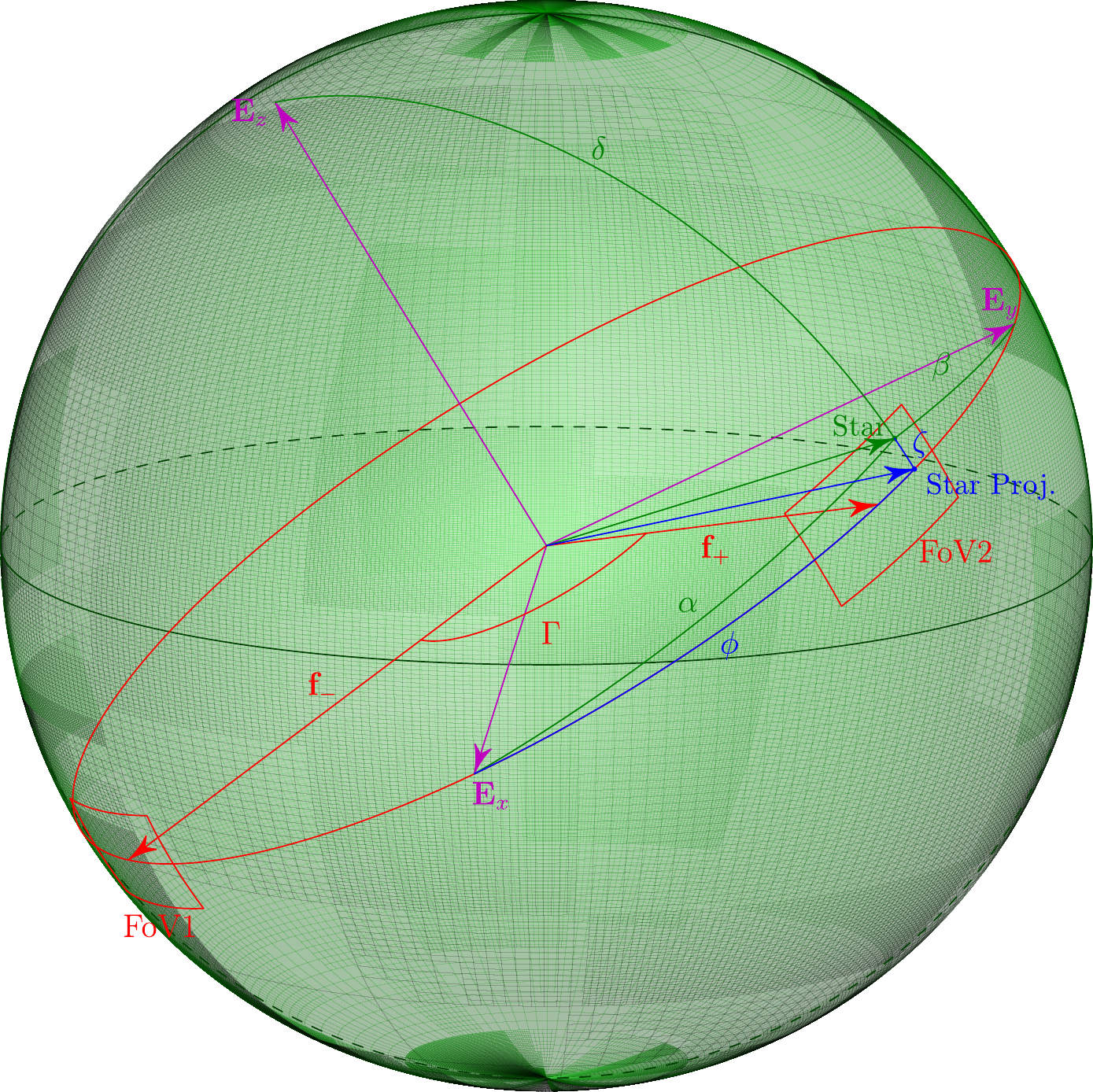}
  \end{center}
  \caption{Fundamental angles in the Gaia reference frame $\mathbf{E}_{\hat a}$. The two Fields of View (FoV) directions are indicated by $f_+$ and $f_-$ while the measured abscissa is given by $n_{(i)}~=~(\alpha,\beta,\gamma)$ (from~\cite{2012SPIE.8451E..3CV}).}
  \label{fig:astroobs_Gaia}
\end{figure}

%%%%%%%%%%%%%%%%%%%%%%%%%%%%%%%%%%
As shown in Fig.~\ref{fig:astroobs_Gaia}, each point of the celestial sphere can be fixed in the reference frame of the Gaia spacecraft by three direction cosines defined as
\be
	n_{(i)} = (\cos \alpha,\cos \beta, \cos \delta) \;, \lb{eq:n_star}
\ee
with $i = 1,2,3$.

%From a geometrical point of view, Gaia measures the abscissa of such a point, $i.e.$ the angle $\phi$ between the $x$-axis 
From a geometrical point of view, Gaia measures the abscissa $\phi$ and
the ordinate $\zeta$, also called along-scan (AL) and across-scan (AC)
coordinate, of such a point. In particular, the AL coordinate is the angle
$\phi$ between the x-axis
of the spacecraft, denoted $E_x$ in Fig.~\ref{fig:astroobs_Gaia}, and the projection of the point along the great circle traced by the $E_x$ and $E_y$ axes, which identifies the instantaneous scanning direction of the satellite.
%in the $x-y$ plane. 
The angle $\phi$ is related to the direction cosines $n_{(i)}$ by the following relations
\bsube \lb{eq:sincosphi}
	\bea 
		\cos\phi &=& \frac{n_{(1)}}{\sqrt{1- n^2_{(3)}}} \; , \lb{eq:cosphi} \\
		\sin \phi &=& \frac{n_{(2)}}{\sqrt{1- n^2_{(3)}}} \; . \lb{eq:sinphi}
	\eea
\esube	
%where the abscissa $n_{(i)}$ is given by Eq.~\eqref{eq:astro} as function of the metric tensor and its derivatives.

As illustrated in~\citet{2016A&A...595A...1G} and \citet{2001A&A...369..339P}, the Gaia spacecraft has two Fields of View (FoV) called $f_+$ and $f_-$ whose pointing directions are separated by a fixed base angle of $106.5^o$ and which are symmetric with respect to the $x$-axis. Since the angular amplitude of each FoV is about $0.5^o$, the abscissae range is fixed in the intervals $53.25 \pm 0.25$ degrees for $f_+$ and $-53.25 \pm 0.25$ degrees for $f_-$.
Therefore, one of Eqs.~\eqref{eq:sincosphi} is enough to determine a univocal correspondence between the value of $\phi$ and that of the direction cosines. 
%The usual choice is $\cos \phi$, so that the director cosines with respect to the $x$ and $z$ axes are sufficient to completely determine the observation.
%Gaia's measurement is $3$ times more sensitive in the along-scan direction than in the across one, so in principle one can only use the $\cos{\phi}$ equation, which is what we have done in this paper; however, since to first order the abscissa measurement is independent on the orientation of the $E_z$ axis, the $\sin{\phi}$ equation should also be used when stellar and attitude parameters are both being reconstructed.
Gaia's measurement is $3$ times more sensitive in the along-scan direction
than in the across one, so, in principle, the $\cos{\phi}$ observation
equation is sufficient to solve the sphere problem, which is what we have
done in this paper; however since to first order the abscissa measurement
is independent from the orientation of the $E_z$ axis, one should also build
an observation equation using the AC measurement $\zeta$ to better
constrain the attitude when the latter is being reconstructed along with
the astrometric parameters.
The abscissa is generally expressed as function of the astrometric parameters (namely, the right ascension $\alpha_*$, declination $\delta_*$, parallax $\varpi_*$, and proper motions $\mu_{\alpha*}$ and $\mu_{\delta*}$) and of the satellite attitude $a_i^{(j)}$, where the index $j$ refers to the time of observation, and $i$ to the number of parameters used to model the attitude. The latter has to be considered unknown since the satellite attitude cannot be determined by other independent measurements at the accuracy required by the mission. For a similar reason Eq.~\eqref{eq:cosphi} also depends on a set of instrument parameters $\{ c_l \}$ to provide a sort of long-term calibration. 
Moreover, when working within the Parametrized Post-Newtonian (PPN) formalism, one should add the parameter $\gamma$ to the unknowns. A better determination of $\gamma$, which measures the amount of curvature induced by the mass-energy on space-time, shall be one of the important scientific contributions of Gaia~\citep{2001A&A...369..339P}.
%As consequence, each one of Gaia observations can be resumed to a non-linear function of these four classes of unknown included in a suitable model of the abscissa $\phi$ as
As consequence, each along-scan observation can be modeled by a non-linear function of these four classes of unknowns, as
\be
	\cos \phi \equiv {\cal F} \Big( \alpha_*,\delta_*, \varpi_*, \mu_{\alpha*}, \mu_{\delta*}, a^{(j)}_1,a^{(j)}_2,... ,c_1,c_2,..., \gamma  \Big) \; .  \lb{eq:obseq}
\ee

\subsection{The GSR approach to the processing of observations} \label{sect:GSRproc}
%%%%%%%%%%%%%%%%%%%%%%%%%%%%%%%%%%
The Gaia mission will perform several billions of observations during its operational years, so that the problem of the so-called ``sphere reconstruction'' translates into the solution of a very large system of equations (up to $10^{10} \times 10^8$ in the case of Gaia).
Solving such a big system of non-linear equations is not feasible, so the observation equations~\eqref{eq:obseq} are linearized about a convenient starting point, $i.e.$ the current best available estimate of the required unknowns. The problem is then converted in a corresponding system of linear equations
\bea \lb{eq:eqphi_lin}
	- \sin \phi \, \Delta \phi &=& \frac{\partial {\cal F} }{\partial \alpha_*} \vert_{\bar \alpha_*} \delta \alpha_* + \frac{\partial {\cal F} }{\partial \delta_*} \vert_{\bar \delta_*} \delta \delta_* +\frac{\partial {\cal F} }{\partial \varpi_*} \vert_{\bar \varpi_*} \delta \varpi_*+ ... \nn \\
		&& +\sum_{i j} \frac{\partial {\cal F} }{\partial a_i^{(j)} } \vert_{\bar a_i^{(j)} } \delta a_i^{(j)} +\sum_{i} \frac{\partial {\cal F} }{\partial c_i} \vert_{\bar c_i} \delta c_i+ ... \nn \\
		&& + \frac{\partial {\cal F} }{\partial \gamma} \vert_{\bar \gamma} \delta \gamma + ... \; ,
\eea
where the unknowns are the corrections $\delta x$ to the a-priori values ($e.g.$, from stellar catalogs) while the derivatives of ${\cal F} $ are the coefficients of the design matrix. The known-terms are then represented by the left-hand side of Eq.~\eqref{eq:eqphi_lin} as
\be \lb{eq:ktdef}
	\sin \phi \, \Delta \phi = \sin (\phi_{calc}) \times (\phi_{obs} - \phi_{calc}) \; ,
\ee
where $\phi_{obs}$ represents the observed abscissa and formally includes the measurement errors so that it can be written as $\phi_{obs} = \phi_{true} + \Delta \phi$, while $\phi_{calc}$ is the computed value given by $\arccos({\cal F} )$ at the starting point of the linearization (generally speaking, the value contained in the astrometric catalog).

The resulting system of equations is quite sparse since each observation refers to a single star among the millions considered in the reconstruction problem (and then only to its astrometric parameters). A similar reasoning is valid for the attitude and calibration parameters, while $\gamma$ is a global parameter in the sense that it appears in each equation of the system.
The number of observations being far larger than the number of unknown parameters, the system is over-determined and can be solved by a least-squares procedure. 
The final goal of the entire procedure is then to get better estimates of all the intervening parameters but for that, we first need to provide an accurate model of the direction cosines of the observation.

%These steps required to modify the following sections of the GSR code, keeping as much as possible its original structure in order to take advantage of the many functionalities already implemented. 
The input of the code are \emph{packets}, each containing observations characterized by: the coordinate time of observation (necessary to retrieve Gaia state vector and the planetary ephemerides at the appropriate epoch), the catalog coordinates of the observed source and all quantities used in the setup of the observation equations for the astrometric problem, which we are going to detail in the following. The values contained in the packet are then updated by the processing of the observation.
For each of the processed observations the following steps are performed:
\begin{itemize}
	\item Load the observations packets and launch the analysis routines;
	\item Call, for each observation, the routines defining all needed quantities for the computation of the the astrometric observable;
	\item Define all needed vectors (star-observer, perturbing body-observer, ...) and tensors (tetrad components, metric, ...); 
	\item Define $\phi_{obs}$ and $\phi_{calc}$, compute the known-terms~\eqref{eq:ktdef};
	\item Compute the coefficients of the linearized observation equation~\eqref{eq:eqphi_lin};
	\item Coefficients and known-terms are associated to an observation and stored in a packet to be then used for the setup of the observation equation and the astrometric solution in a least-squares sense.
\end{itemize}

The GSR software has been conceived to be as neutral as possible with respect to the astrometric model. In particular, as long as the latter agrees with some given Input/Output specifications, one can plug-in any relativistic model, which is seen as a ``black box'' by the pipeline. 
In the current version of GSR (GSR2, see section~\ref{sect:GSRTTFres}) the direction cosines are provided using a relativistic model from the RAMOD family (more precisely, the version actually implemented in the operational infrastructure at DPCT is RAMOD@GSR2,~\cite{AVDemonRun17})
%Currently, the direction cosines are provided using a relativistic model from the RAMOD family (more precisely, the version actually implemented is PPN-RAMOD~\citep{2003A&A...399..337V})
, but the modular structure of the software makes it easy to produce a GSR-compatible implementation of the TTF astrometric model, which we denote as GSR-TTF.

%%%%%%%%%%%%%%%%%%%%%%%%%%%%%%%%%%
\section{TTF implementation in GSR} \label{sect:GSRTTF}
%%%%%%%%%%%%%%%%%%%%%%%%%%%%%%%%%%
The goal of this work is to setup the processing of astrometric observations (eventually made by Gaia) using the TTF formalism. To do it, we implement our model in the GSR software developed at Turin Observatory and we use it to generate a series of simulated observations.
%A brief overview of our activity on the GSR code is given in Fig.~\ref{fig:GSRschema}. We mainly focused on the computation of the Gaia observable and on writing the linearized observation equations necessary to evaluate the astrometric coordinates of the observed star from a series of observations.
The result is a ``GSR-TTF plug-in'' which enables the pipeline to reduce the sphere with another model, which is actually more accurate than the RAMOD@GSR2 one. Actually, the TTF model can be implemented at several different level of accuracy. We decided to limit the present one to a $(v/c)^3$ version at the same level of GREM, thus neglecting all the complications coming from higher-orders of $v/c$ and to non-spherically symmetric gravity fields, which are deemed negligible for the specific problem of the global sphere reconstruction.

The implementation of the astrometric observable in GSR concerns the development of both sides of Eq.~\eqref{eq:eqphi_lin}, as we detail in this section. 
%In this section we will focus on building and testing the known terms~\eqref{eq:ktdef}.
%In this sections, we will detail the procedure followed to implement the astrometric model presented in this thesis into the GSR software as well as the tests performed on our results. 
 
%%%%%%%%%%%%%%%%%%%%%%%%%%%%%%%
\subsection{Setup of the observation abscissae} \lb{sec:abscissaesimul}

Let us show the procedure followed to build the abscissa $\phi$ using our astrometric model.
First, following the definition of astrometric observable in the tetrad formalism~\citep{1991ercm.book.....B} given by~\citet{2014PhRvD..89f4045H}, we define the direction cosines appearing in Eq.~\eqref{eq:cosphi}, taken at the observation point $x_B$ as
\begin{equation} \lb{eq:nGSR}
	n^{(i)}_B=-\frac{E^0_{(i)}+E^j_{(i)}\hat k_j}{E^0_{(0)}+E^j_{(0)}\hat k_j } \Bigg\vert_{x_B} \; ,
\end{equation}
where we shall choose the direction triple $\bm \hat k_i$ (defining the barycentric direction of light) and the tetrad $E^\alpha_{(\beta)}$ ($i.e.$ the transformation matrix to the reference system comoving with the observer) according to the accuracy required by the Gaia mission.

%%%%

The relations between the Time Transfer Functions and the wave vectors $k^{\mu}=dx^{\mu}/d\lambda$ at reception have been derived in \citet{2004CQGra..21.4463L} as
%\begin{subequations} \lb{eq:k}
\be
\left(\widehat{k}_i\right)_B = \left(\frac{k_i}{k_0}\right)_B =
-c \, \frac{\partial {\cal T}_{e}}{\partial x^{i}_{B}} \, = \,
- c \, \frac{\partial  {\cal T}_{r}}{\partial x^{i}_{B}}
\left[1 - \frac{\partial  {\cal T}_{r}} {\partial t_B}\right]^{-1} \; .   \label{eq:k} 
%& & \nonumber \\
%& &\frac{(k_{0})_B}{(k_{0})_A} =  \left[1 + 
%\frac{\partial  {\cal T}_{e}}{\partial t_{A}}\right]^{-1} \, 
%= \, 1 -
%\frac{\partial  {\cal T}_{r}}{\partial t_{B}}  \quad . \label{2d3}
\ee
%\end{subequations}
\noindent   

We put $x_A = (ct_A, \bx_A)$ the event of emission ${\cal A}$ and $x_B = (ct_B, \bx_B)$ the event of reception ${\cal B}$ of a light signal. Moreover, we define ${\cal T}_{e}$ and ${\cal T}_{r}$ as two distinct (coordinate) time transfer functions defined as  
\begin{equation}  \label{ttf}
t_B - t_A = {\mathcal T}_e(t_A, \bx_A, \bx_B) = {\mathcal T}_r(t_B, \bx_A, \bx_B) \, .
\end{equation}
where $\mathcal T_e$ and $\mathcal T_r$ are evaluated at the event of emission ${\cal A}$ and at  the event of reception ${\cal B}$ respectively. For astrometric applications, we are only interested in $\mathcal T_r$.
%ADD A MORE DETAILED JUSTIFICATION FOR T_R ONLY?

For applications in the Solar System, we can work in the weak field approximation so that we can write
\begin{equation}
g_{\mu\nu}=\eta_{\mu\nu}+h_{\mu\nu}\, ,
\end{equation}
with $\eta_{\mu \nu}=diag(-1,+1,+1,+1)$ a Minkowskian background and $h_{\mu \nu}$ a small perturbation. Moreover, we shall consider only weak gravitational fields generated
by self-gravitating extended bodies within the slow-motion, post-Newtonian approximation. So, we assume that the potentials $h_{\mu \nu}$ may be expanded as
\bea \lb{1n}
%& & h_{00}=h_{00}^{(2)}+h_{00}^{(4)}+{\cal O}\left(\frac{1}{c^6}\right) \nonumber \, , \\
& & h_{00}=h_{00}^{(2)}+{\cal O}\left(\frac{1}{c^4}\right) \nonumber \, , \\
& & h_{0i}=h_{0i}^{(3)}+{\cal O}\left(\frac{1}{c^5}\right) \, , \\ 
& & h_{ij}=h_{ij}^{(2)}+{\cal O}\left(\frac{1}{c^4}\right) \, , \nonumber
\eea
where $h_{\alpha\beta}^{(2n)}\sim\left(U/c^2\right)^{n}$, $h_{\alpha\beta}^{(2n+1)}\sim-\left(U/c^2\right)^{n}(v/c)$ and $U$ is the Newtonian potential.

Under these hypothesis, the expression of the time transfer functions $\mathcal{T}_{e/r}$ in the PPN approximation has been given in~\cite{2002PhRvD..66b4045L,2008ASSL..349..153T} as
\be \lb{ttf_order}
	{\cal T}_{	r}(\bx_A, t_B, \bx_B)=\frac{R_{AB}}{c}+ \frac{1}{c} \Delta_{r}(\bx_A, t_B,\bx_B) + {\cal O}(c^{-5}) \; , 
\ee
where $\Delta_{r}$ is defined as
\be
	\Delta_{r} = \frac{1}{2}R_{AB}\int_{0}^{1}\left[h^{(2)}_{00} + \frac{2}{c} N_{AB}^{i} h^{(3)}_{0i} + N_{AB}^{i} N_{AB}^{j} h^{(2)}_{ij}\right]_{z^\alpha_{-}(\lambda)} d\lambda  \lb{Tr1wIAUs}
\ee
and the integrals are taken along the Minkowskian paths $z^\alpha_{-}(\lambda) = (x^0_B - \lambda R_{AB}, x^i_B - \lambda R^i_{AB})$ and we define $R^i_{AB}=x^i_B-x^i_A$, $R_{AB}=\vert {R^i_{AB}} \vert$ and $N^i=\dfrac{R_{AB}^i}{R_{AB}}$.

The derivatives needed in the definition of the wave vectors in Eq.~\eqref{eq:k} can then be computed as (see, $e.g.$, \cite{2014CQGra..31a5021B,2014PhRvD..89f4045H,2014PhRvD..90h4020H})
\begin{subequations}
\be
	\frac{\partial {\cal T}_{r}}{\partial x^i_{A/B}}= \mp \frac{N^i_{AB}}{c} + \frac{1}{c}\frac{\partial \Delta_{r}}{\partial x^i_{A/B}} + {\cal O}(c^{-5})
\ee
and
\be
	\frac{\partial {\cal T}_{r}}{\partial t_{B}}= \frac{1}{c}\frac{\partial \Delta_{r}}{\partial t_B} + {\cal O}(c^{-5}) \; ,
\ee
\end{subequations}
where
\begin{subequations}
\be \lb{eq:dDr1PM}
	\frac{\partial \Delta_r }{\partial x^i_B} = -\frac{1}{2} \int_0^1 \left[  R^i_{AB} \lambda m_{,0} - R_{AB} (1-\lambda) m_{,i} - \tilde h_i  \right]_{z_- (\lambda)} d\lambda
\ee
and
\be
	\frac{\partial \Delta_r }{\partial t_B} = c \frac{R_{AB}}{2} \int_0^1 \left[ m_{,0}  \right]_{z_-(\lambda)} d\lambda \; .
\ee
\end{subequations}
Moreover, we defined
\bea
	m_{,\alpha} &\equiv& h_{00,\alpha} + 2 N_{AB}^k h_{0k,\alpha} + N_{AB}^j N_{AB}^k h_{jk,\alpha} \nn\\
	\tilde h_i &\equiv& N_{AB}^i h_{00} - N_{AB}^i N_{AB}^j N_{AB}^k h_{jk} + 2 h_{0i} + 2 N_{AB}^j h_{ij} \; . \nn
\eea

For most observations, at Gaia accuracy we shall only consider the PN gravitational potentials of all Solar System bodies as point masses. Quadrupole terms might be significant for a limited number of observations grazing giant planets ($e.g.$, closer than $152^{\prime\prime}$ from Jupiter's limb at $1~\mu$as accuracy~\citep{2003AJ....125.1580K,2006CQGra..23.4853C}), hence we are including them in our implementation.

Also, it has been shown \citep{2003A&A...399..337V} that thanks to the satellite's high measurement accuracy the Gaia sphere reconstruction is sensitive to variations of the $\gamma$ parameter of the PPN formalism up to the $10^{-6}-10^{-7}$ level. This processing can thus be used as a tool to improve our knowledge of such parameter, which is currently set at the $10^{-5}$ level \citep{2014LRR....17....4W}. With this application in mind $\gamma$ can be straightforwardly introduced by using the PPN expression of the above metric coefficients $h_{\alpha\beta}$, where
\bea
% 	h_{00} & = & \frac{2 G}{c^2} \sum_P \frac{{\cal M}_P}{r_P (t)} \; , \\
%  	h_{0i} & = & -(1+\gamma) \frac{2 G}{c^2} \sum_P \frac{{\cal M}_P}{r_P (t)}\frac{v_P (t)}{c} \; , \\
%  	h_{ij} & = & \delta_{ij} \, \gamma \, h_{00} \; .
	h_{00} & = & \sum_P \frac{2 w_P(t)}{c^2} \; , \\
 	h_{0i} & = & -(1+\gamma) \frac{2}{c^3} \sum_P w_P(t) v_P (t) \; , \\
 	h_{ij} & = & \delta_{ij} \, \gamma \, h_{00} \;
\eea
and $w_P(t)$ is the time-dependent gravitational potential of perturbing body $P$.
%Moreover, it has to be stressed that the presence of time-dependent positions $r_P (t)$ and velocities $v_P (t)$ of the perturbing bodies 
This time-dependence would in principle require either to express $w$ as a retarded potential, or to compute a full integral of the geodesic. 
However, it has been shown in~\cite{2003AJ....125.1580K} and confirmed by our analysis in~\cite{2014CQGra..31a5021B}, that at the Gaia accuracy the effects of a time-dependent perturbation due to a moving body can be taken into account simply by computing the positions of the gravitating bodies at the retarded moment of time
\be \lb{eq:rettimes}
	t_C = t_B - c^{-1} \abs{\bx_B - \bx_P (t_B)} + {\cal O} (c^{-2}) \; ,
\ee
where $t_B$ is the coordinate time of observation and $\bx_P (t_B)$ is the position of body $P$ at $t_B$, and that the ``gravitomagnetic'' contributions proportional to $v_P (t)$ can be neglected.

Thus, using such constant value $\bx_P = \bx_P (t_C)$ in our computations yields to the following expression of the PN terms of the metric tensor:
\bsube \lb{eq:staticretmetric}
	\bea 
% 		h_{00} &=& \frac{2 G}{c^2} \sum_P \frac{{\cal M}_P}{r_P (t_C)} \; , \\
%  		h_{0i} &=& 0 \; , \\
% 		h_{ij} &=& \delta_{ij} \, \gamma \, h_{00} \; ,
		h_{00} & = & \sum_P \frac{2 w_P(t_C)}{c^2} \; , \\
 		h_{0i} & = & 0 \; , \\
 		h_{ij} & = & \delta_{ij} \, \gamma \, h_{00} \; ,
	\eea
\esube
where $\bm r_P=\bx_B - \bx_P (t_C) - \lambda \rab \bnab$.

%(\red {- I have also implemented the $J_2$ using Le Poncin 2004 and an average z-axis for the KT ... to be tested on some very close obs!}-)
%
%\red{We neglect here the multipole terms since their implementation in the code would be quite cumbersome and their influence would be significative only for observations grazing the giant planets ($1 \; \mu as$ at $152"$ from Jupiter).}

%%%%%%%%%%%%%%%%%%%%%%%%%%%%%
Let us assume that the smallest sphere containing the body has a radius equal to the equatorial radius $r_e$ of the body and that the segment joining $\bx_A$ and $\bx_B$ is outside this sphere. At any point $\bx$ such that $r \geq r_{e}$, the gravitational potentials $w$ and $\bm w$ are then given (for each body $P$) by the multipole expansion~\cite{1980RvMP...52..299T,1997JMP....38.2587K,2002PhRvD..66b4045L}
\be \lb{eq:gravpotmultip}
w = \frac{G {\cal M} }{r}\left[1 - \sum_{n=2}^{\infty} J_n \left(\frac{r_e}{r}\right)^n P_n \left(\frac{\bk .\bx}{r}\right)\right] \quad \mbox{and} \quad \bm w =0 \; ,
\ee
where $\bk$ denotes the unit vector along the $x^3$-axis, the $P_{n}$ are the Legendre polynomials, ${\cal M}$ is the mass of the body and the coefficients $J_n$ are the mass multipole moments. 

%%%%%%%%%%%%%%%%

Under these assumptions, the monopolar and quadrupolar terms of the direction triple to be used in Eq.~\eqref{eq:nGSR} are~\citep{2008PhRvD..77d4029L, 2013sf2a.conf..155B}
\begin{widetext}
\bsube \lb{eq:deflmultip}
\bea \lb{eq:deflGSR}
(\hat k_i^B)(\bx_A,\bx_B) &=&  \nabi - (\gamma+1) \sum_P \frac{G {\cal M}_P}{c^2 R_{PB}}\, \frac{1}{1 + \bN_{PA} \cdot \bN_{PB}}  \times \left[\frac{R_{AB}}{R_{PA}}\bN_{PB} -\left(1 + \frac{R_{PB}}{R_{PA}}\right)\bN_{AB}\right]  \; , \lb{M0}
\eea
\bea \lb{eq:deflGSR_quadr}
(\hat k_i^B)_{J_2}(\bx_A,\bx_B) &=& (\gamma +1)\frac{GM}{c^2} J_2 r_{e}^2 \Bigg\lbrace - \Big[ \bk \cdot (\bN_{PA} + \bN_{PB}) \Big]^2 \left[\frac{\bN_{PB} - \bN_{AB}}{(R_{PA}+R_{PB}-R_{AB})^{3}}- \frac{\bN_{PB} + \bN_{AB}}{(R_{PA}+R_{PB}+R_{AB})^{3}}\right] + \lb{M2} \nonumber \\
&& \nn \\
& &\frac{1}{2}\left[\frac{1-(\bk \cdot \bN_{PA})^2}{R_{PA}} + \frac{1-(\bk \cdot \bN_{PB})^2}{R_{PB}}\right] \left[\frac{\bN_{PB} - \bN_{AB}}{(R_{PA}+R_{PB}-R_{AB})^{2}} - \frac{\bN_{PB} + \bN_{AB}}{(R_{PA}+R_{PB}+R_{AB})^{2}}\right]  + \nonumber \\ 
& & \frac{1}{R_{PB}^3}\frac{(R_{PA} + R_{PB})R_{AB}}{R_{PA}^{2}}\frac{\bk . (\bN_{PA} + \bN_{PB})}{(1 + \bN_{PA} \cdot \bN_{PB})^2} \Big[ \bk - (\bk \cdot \bN_{PB})\bN_{PB} \Big] + \\ 
& & \frac{1}{2 R_{PB}^3}\frac{R_{AB}}{R_{PA}}\frac{2(\bk \cdot \bN_{PB})\bk + \left[1 - 3(\bk \cdot \bN_{PB})^2\right]\bN_{PB}}{1 + \bN_{PA} \cdot \bN_{PB}}\Bigg\rbrace \; \nn ,   
\eea
\esube
\end{widetext}
respectively, and where we used
\bsube \lb{eq:defkGSR}
\bea 
&&\bm{R}_{PX}=\bx_X - \bx_P(t_{C}) \quad ,  \\
&&  R_{PX}=|\bm{R}_{PX}| \qquad , \qquad \bm{N}_{PX} = \frac{\bm{R}_{PX}}{R_{PX}} \; .  %\\
%&&\bm{R}_{AB}=\bx_B-\bx_A,~~R_{AB}=|\bm{R}_{AB}|,~~\bm{N}_{AB} = \frac{\bm{R}_{AB}}{R_{AB}} \; . 
\eea
\esube
An expression for the multipolar terms of higher order is given in~\cite{2008PhRvD..77d4029L} but it is not relevant for this study. 

\vspace{0.5cm}
%The contribution of the mass multipole moments to the deflection of light has to be taken into account in astrometric missions of high precision such as Gaia for light rays grazing the deflecting body (see Fig.~\ref{Table-gravitational-deflection} and discussion in~\cite{2003AJ....125.1580K,2008PhRvD..77d4029L}). It shall then be included in an astrometric model aiming at interpreting Gaia's observations.
%In this sense, Eqs.~\eqref{eq:Deltamultip}-\eqref{eq:deflmultip} show that there would be no theoretical limitation to include the quadrupole light deflection term in an astrometric model based on the TTF formalism (see chapter~\ref{chap7}).

%\subsection{Description of Gaia attitude and frame transformations} \label{sect:tetrad}
%%%%%%%%%%%%%%%%%%%%%%%%%%%%%%%%%%
The direction triple $(\hat k_i^B)$, defined as the sum of Eqs.~\eqref{eq:deflGSR}-\eqref{eq:deflGSR_quadr}, gives the direction of the light ray at reception in the local barycentric frame. Following Eq.~\eqref{eq:nGSR}, we hence need to project this vector in the observer reference frame.
Concerning this point, we have used the tetrad defined 
%the simplest choice is to use the tetrad presented
in~\citet{2010A&A...509A..37C} and already used in GSR, evaluated with the same metric~\eqref{eq:staticretmetric}.

The satellite reference frame defining the transformation $E^\beta_{\hat a}$ is obtained by successive transformations of the local BCRS tetrad~\citep{2003CQGra..20.4695B} 
\bea
	\lambda^\alpha_{\hat a} &=& h_{0 \alpha} \delta_0^\alpha + \left( 1- \frac{h_{aa}}{2} \right)\delta^\alpha_a + {\cal O} (h^2) \; , \\\nonumber
     &=&  h_{0 \alpha} \delta_0^\alpha + \left( 1- \gamma \frac{h_{00}}{2} \right)\delta^\alpha_a + {\cal O} (h^2)
\eea 
namely the tetrad obtained from the BCRS by shifting the origin to the instantaneous barycenter of the satellite. This tetrad, and in particular the triad of four-vectors $\bm \lambda_{\hat a}$ defined as the spatial part of $\lambda^\alpha_{\hat a}$, is ``boosted'' to the satellite rest-frame by means of an instantaneous Lorentz transformation identified by the four-velocity of the observer $u_s$ with respect to the local BCRS. The boosted tetrad $\bm \tilde{\lambda}_{\hat a}$, namely \citep{2010A&A...509A..37C}
\be \lb{eq.bs.tetrad}
   \tilde{\lambda}_{\hat a}^{\alpha}=\lambda^\alpha_{\hat a}+
   \left(1+3\gamma\frac{h_{00}}{2}+\frac{1}{2}\frac{v^2}{c^2}\right)\delta^\alpha_0\frac{v^a}{c}+\frac{1}{2}\frac{v^i}{c}\delta^\alpha_i\frac{v^a}{c}
\ee
obtained in this way represents a reference system whose origin is comoving with the barycenter of the satellite and whose spatial axes are kinematically non-rotating with respect to the BCRS. The Gaia attitude frame is finally obtained with a rotation of these spatial axes. Since the transformation is Euclidean, the rotation matrix $A$ is exactly the attitude matrix of the satellite.

The triad resulting from these transformations, detailed in~\cite{2010A&A...509A..37C}, establishes 
%Gaia attitude triad given by
the Gaia relativistic attitude triad and is related to the euclidean attitude parameters by
\be \lb{eq.obsRAMOD}
	\begin{pmatrix}
		E_{\hat{1}}\\
		E_{\hat{2}}\\
		E_{\hat{3}}
	\end{pmatrix}=A
	\begin{pmatrix}
		\tilde{\lambda}_{\hat{1}}\\
		\tilde{\lambda}_{\hat{2}}\\
		\tilde{\lambda}_{\hat{3}}
	\end{pmatrix}.
\ee

The temporal axis of the tetrad has to be transformed alike to obtain a complete reference system attached to the observer. This result is naturally achieved by writing the transformation between the barycentric coordinate time and the observer's proper time.

Finally, introducing Eq.~\eqref{eq:nGSR} into Eq.~\eqref{eq:cosphi} completes the implementation in the observation abscissa of GSR-TTF.

%%%%%%%%%%%%%%%%%%%%%%%%%%%%%%%
\subsection{Setup of the astrometric coefficients} \lb{sec:partialscompastro}

%At this preliminary phase of the implementation, we content ourselves of estimating the astrometric coordinates $\alpha_*$ and $\delta_*$ so that all other parameters are assumed to be known and we consider the following simplified version of Eq.~\eqref{eq:eqphi_lin}
%\be \lb{eq:eqphi_linsimple}
%	- \sin \phi_{calc} d\phi = \frac{\partial {\cal F} }{\partial \alpha_*} \Bigg \vert_{\bar \alpha_*} \delta \alpha_* + \frac{\partial {\cal F} }{\partial \delta_*} \Bigg\vert_{\bar \delta_*} \delta \delta_*  \; . 
%\ee

In order to setup the observation equation for the improvement of the a priori catalog values by a least-square procedure, we define the partial derivatives of Eq.~\eqref{eq:obseq} w.r.t. the astrometric parameters as in Eq.~\eqref{eq:eqphi_lin}.
The coefficients can be computed analytically since the function ${\cal F} = \cos \phi_{calc}$, defined by Eq.~\eqref{eq:cosphi}, is known. By defining $n_{(i)}=\cos{\psi_{(i)}}$, the coefficients of Eq.~\eqref{eq:cosphi} can be computed analytically as
%These quantities were implemented in the {\em CommonTermsRod.java} as well as the astrometric coefficients appearing at the second member of Eq.~\eqref{eq:eqphi_lin} given by
\be
	\frac{\partial \cos \phi}{\partial v} = \frac{\partial_v (\cos\psi_{(1)})}{\sqrt{1-\cos^2\psi_{(3)}}} + \frac{\cos \psi_{(1)} \cos \psi_{(3)} \partial_v (\cos\psi_{(3)})}{(1-\cos^2\psi_{(3)})^{3/2}} \; \lb{eq:dcosphidv},
\ee
where $v$ are the so called astrometric parameters characterizing the coordinates of a star at a moment in time: the parallax $\varpi_*$, right ascension $\alpha_*$, declination $\delta_*$, and the proper motions $\mu_{\alpha_*}$ and $\mu_{\delta_*}$ in perpendicular directions.
The partial derivatives appearing therein are given by
\begin{widetext}
\be \lb{eq:dercosdir}
	\qquad \qquad \qquad \qquad \quad 
\partial_v \cos \psi \equiv \frac{\partial \cos\psi_{(i)}}{\partial v}  
			= - \frac{E^j_{(i)} \partial_v \hat k_j (E^0_{(0)} + E^l_{(0)} \hat k_l) - (E^0_{(i)} + E^j_{(i)} \hat k_j) E^l_{(0)} \partial_v \hat k_l}{ (E^0_{(0)} + E^l_{(0)} \hat k_l)^2 } \quad ,   \\
\ee
%\end{widetext}
with
%\begin{widetext}
\bea
	\frac{\partial \hat k_i}{\partial v} &=& \partial_v \nabi - (\gamma + 1) \sum_P \frac{G {\cal M}_P}{c^2 \rpb} \frac{1}{1+ \bnpa \cdot \bnpb} \Bigg\{  - \frac{\partial_v \bnpa \cdot \bnpb}{1+ \bnpa \cdot \bnpb}  \times \Bigg[ \nabi \frac{\rab}{\rpa} - \nabi \Bigg(1 + \frac{\rpb}{\rpa}\Bigg) \Bigg] + \nn \\
		&& \frac{N_{PB}^i}{\rpa^2} \Big(\rpa \partial_v \rab - \rab \partial_v \rpa \Big)  - \partial_v \nabi \Bigg(1+\frac{\rpb}{\rpa} \Bigg) + \nabi \rpb \frac{\partial_v \rpa}{\rpa^2} \Bigg\} \; . 
\eea
\end{widetext}
Moreover, we used the definitions~\eqref{eq:defkGSR} along with
\begin{subequations}
\bea
	\partial_v \nabi &=& - \frac{(\bnab \times \partial_v \bx_A \times \bnab)^i}{\rab} \; , \\
	\partial_v \rab &=& - \frac{\bm \rab \cdot \partial_v \bx_A}{\rab} \; , \\
	\partial_v \rpa &=& \frac{\brpa \cdot \partial_v \bx_A}{\rpa} \; , \\
	\partial_v \bnpa &=& \frac{1}{\rpa} \Big[ \partial_v \bx_A - \bnpa (\bnpa \cdot \partial_v \bx_A) \Big] \; .
\eea
\end{subequations}
The expression in terms of the astrometric parameters can be easily obtained by considering that $\bx_A(t)$ can be treated like a Euclidean vector, so that $\bx_A(t)=\left(r_A\cos\alpha(t)\cos\delta(t),r_A\sin\alpha(t)\cos\delta(t),r_A\sin\delta(t)\right)$, where $r_A = \norm{\bx_A}$ is the barycentric distance of the star, $\varpi \equiv 1\;UA / r_A$. Finally, the angular coordinates $\alpha(t)$ and $\delta(t)$ at a given time $t$ can be expressed as functions of the catalog positions and proper motion with, e.g., $\alpha(t) = \alpha_{t_0} + \mu_{\alpha_*} dt$, $dt$ being the interval of time between $t$ and the catalog reference time $t_0$. We can thus write
\begin{subequations}
\bea
	\partial_{\varpi_*} x^i_A &=& - \frac{1 \; UA}{\varpi_A} \times \bn_A\; , \\
	\partial_{\alpha_*} x^i_A &=& (- r_A \sin \alpha \cos \delta, r_A \cos \alpha \cos \delta, 0) \; , \\
	\partial_{\delta_*} x^i_A &=& (- r_A \cos \alpha \sin \delta, - r_A \sin \alpha \sin \delta, r_A \cos \delta) \; , \\
	\partial_{\mu_{\alpha_*}} x^i_A &=& dt \times  \partial_{\alpha_*} x^i_A\; , \\
	\partial_{\mu_{\delta_*}} x^i_A &=& dt \times  \partial_{\delta_*} x^i_A \; .
\eea
\end{subequations}
%The problem of solving Eq.~\eqref{eq:obseq} is well posed if we consider the minimisation of the total residual
%\be
%		\sum_i \norm{ \cos \phi_i - {\cal F}_i (\alpha_*,\delta_*) }  \; ,
%\ee 
%which can be linearized around a convenient value to obtain 
The system is then solved following a classical least-squares method.

%%%%%%%%%%%%%%%%%%%%%%%%%%%%%%%
\subsection{Setup of the global PPN parameter $\gamma$} \lb{sec:partialscompgamma}

Following a similar approach, we computed and implemented the partial derivatives of the astrometric observable w.r.t. the global PPN parameter $\gamma$, which can be used to test General Relativity using light propagation measurements.

Considering Eq.~\eqref{eq:dcosphidv}, we have that both the tetrad $E^ \alpha_\beta$ and the direction triple $\hat k_B$ depend on $\gamma$ through the metric~\eqref{eq:staticretmetric}.
Hence we get
%\newpage

\begin{widetext}
\be \lb{eq:dercosdir}
	%\qquad \qquad \qquad \qquad \quad 
\partial_\gamma \cos \psi \equiv \frac{\partial \cos\psi_{(i)}}{\partial \gamma}  
			= - \frac{ ( E^j_{(i)} \partial_\gamma \hat k_j + \partial_\gamma E^j_{(i)}  \hat k_j  ) (E^0_{(0)} + E^l_{(0)} \hat k_l) - (E^0_{(i)} + E^j_{(i)} \hat k_j) E^l_{(0)} \partial_\gamma \hat k_l}{ (E^0_{(0)} + E^l_{(0)} \hat k_l)^2 } \quad ,   \\
\ee
%\end{widetext}
and, with Eqs.~\eqref{eq:deflmultip},  \eqref{eq.bs.tetrad} and \eqref{eq.obsRAMOD},
%\begin{widetext}
\bea
	\frac{\partial \hat k_i}{\partial \gamma} &=& \frac{(\hat k_i^B) + (\hat k_i^B)_{J_2} }{(\gamma + 1)} \qquad and \qquad  \partial_\gamma E^0_{(i)} = \partial_\gamma E^\alpha_{(0)} = 0 \; .
\eea
\end{widetext}

%%%%%%%%%%%%%%%%%%%%%%%%%%%%%%%%%%
\section{GSR-TTF : results with Gaia simulated observations} \label{sect:GSRTTFres}
%%%%%%%%%%%%%%%%%%%%%%%%%%%%%%%%%%
Our implementation is tested by using observations of the Gaia satellite as simulated by DPAC using the GREM model. In particular, the dataset employed is RDS-7-F~\citep{2010LL....XL-020-01}, a full  5-yr simulation of Gaia observations of 2 million stars, approximately 1 million of which are Primary astrometric sources in the sense discussed in~\cite{2012A&A...538A..78L}. Besides the noise-free measured GREM abscissa $\eta$ (see Eq.~\eqref{eq:phiGREM} below), the data set provides: stellar coordinates, observation epoch, instantaneous satellite attitude, and the ephemerides of the Solar System. These data allow us to compute the abscissa $\phi$ from a given model.
As for the observation error, this can be customized. In our experiments all stars have Gaia magnitude $G=14$, and at such brightness the expected single observation error is
$\sigma_0 = 125\;\mu$as. 

%It is also possible, for each observation, to compute the true along-scan (AL) field angle $\eta$. Then, by definition, the abscissa of Eq.~\eqref{eq:cosphi} can be expressed as
%\be
%	\phi=f\frac{\Gamma}{2}+\eta,
%	\lb{eq:phiGREM}
%\ee
%where $\Gamma$, as shown in figure~\ref{fig:astroobs_Gaia}, is the \emph{basic angle}\/, that is the angle between the axes of the preceeding and following FoV, and $f=\pm1$ with sign determined by the FoV in which the star is observed. In particular $f$ takes the minus sign in the case of $\mathbf{f}_-$, and the positive one for $\mathbf{f}_+$.

The TTF model is therefore tested in two steps: by comparing the abscissae~\eqref{eq:cosphi} computed by TTF and GREM, and by trying some reconstructions of the global astrometric sphere.

%First, we compute the observation abscissa~\eqref{eq:cosphi} for a chosen lapse of time and we compare our results with those of GREM. We thus validate our implementation against GREM by checking the residual differences between this value and the one obtained from the TTF model.
%
%In a second time, we focus on the implementation of the linearized observation equation~\eqref{eq:eqphi_lin} and on testing the procedure of reconstruction of the celestial sphere.

%This analysis gives a global overview and a first validation of how our model can be applied to the complex task of processing the observations of a mission of space astrometry.

The latter had to be restricted to relatively small subsamples of stars and observations in RDS-7-F. This is because the test \& development section of the GSR system at DPCT, the environment for running experiments with simulated data, is allocated only limited resources. 

%We shall now compare our results to those of PPN-RAMOD (actually implemented in GSR) and of GREM (the model implemented in AGIS).

\subsection{Numerical comparison of computed observables} \label{sect:numComp}
%%%%%%%%%%%%%%%%%%%%%%%%%%%%%%%%%%
As anticipated above, we first focused on the comparison of the abscissa $\phi$. Using the implementation described in Section~\ref{sec:abscissaesimul}, we apply Eq.~\eqref{eq:cosphi} over a set of observations from a chosen day of the given dataset using the TTF model, thus obtaining $\phi_\mathrm{TTF}$. The same data are used with the tools provided by the DPAC code, which implements the GREM model, to obtain $\phi_\mathrm{GREM}$. Finally, the residuals $\phi_\mathrm{TTF}-\phi_\mathrm{GREM}$ are computed.

\begin{figure}[hbt]
\begin{center}
	\includegraphics[width=0.48\textwidth]{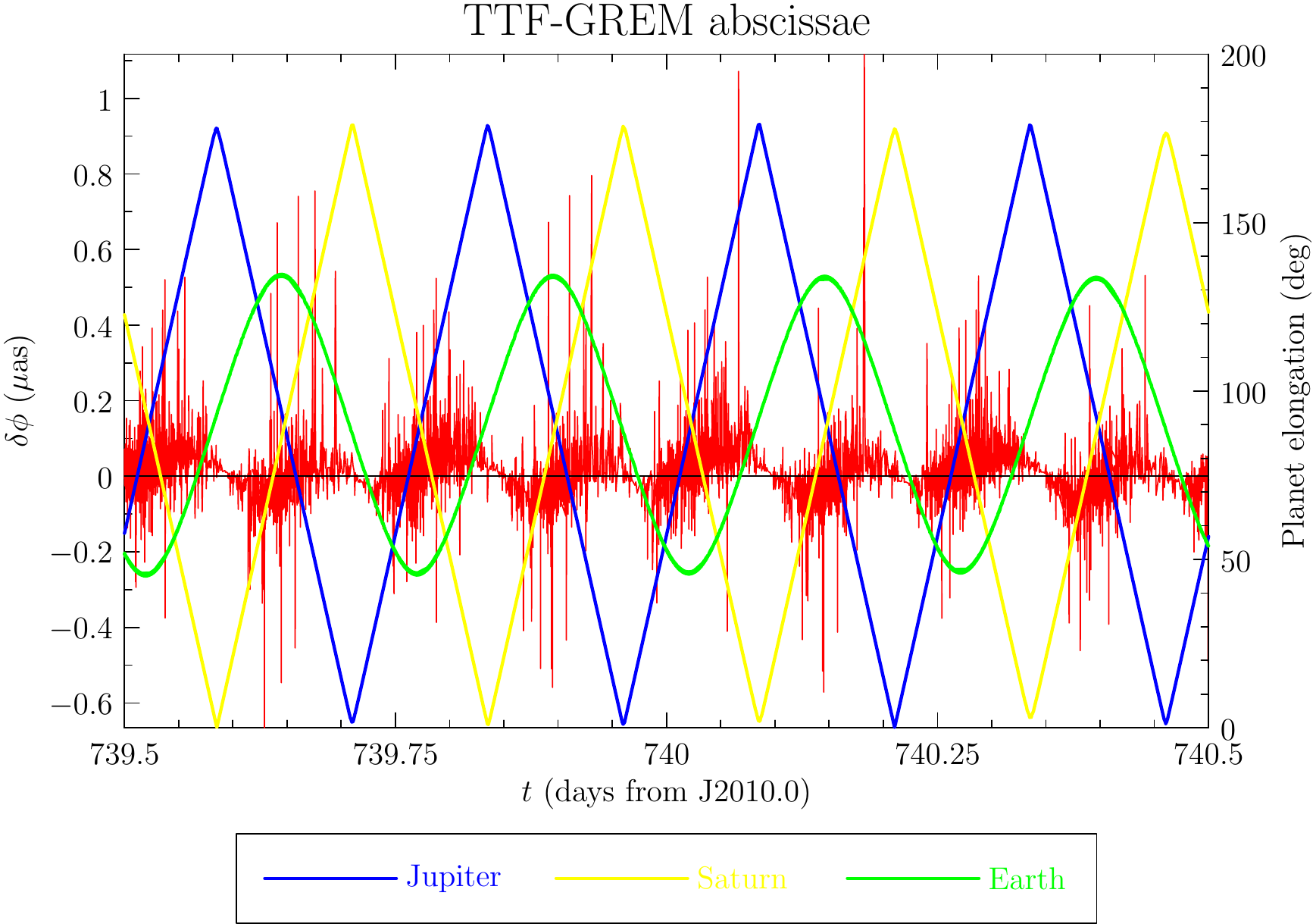}
%\\
% 		\includegraphics[width=0.5\textwidth]{GSR/Series4_Elements81952RAMOD_VS_TTF_AndJupiterDist_NSL703_TTFretTime.jpg}
\end{center}
\caption{Difference between the abscissae resulting from the GSR-TTF implementation and from GREM implementation in the ``GaiaTools". The left axis indicates the difference in $\mu as$ between the two models~-~represented by the red plot; the right axis indicates the distance in degrees between a given planet and the observation - the blue, green and yellow plot representing Jupiter, Saturn and the Earth, respectively.}
\label{fig:confronto1day_1}
\end{figure}

The results are illustrated in Fig.~\ref{fig:confronto1day_1}
% (produced using the Gaia-tools provided by the Gaia DPAC)
, where the models are compared. The left axis marks the difference in $\mu as$ between the two models~-~represented by the red plot~-~while the right axis is the angular distance in degrees between a given planet and the observation, namely its elongation. The blue, yellow and green lines represent Jupiter, Saturn and the Earth, respectively. 
In particular, the periodic oscillation of the distance planet-observation illustrated in the plots is due to the Gaia scanning law~\citep{2010IAUS..261..331D} setting a rotation period of approximately $6\; h$.

As expected from the analytical comparisons in~\citet{2013sf2a.conf..155B} and \citet{2014CQGra..31a5021B}, the plot shows that the differences between the abscissae computed with the two models are generally well below $\pm 1 \; \mu as$.
The remaining sub-$\mu as$ signal can be attributed to a different accounting of the retarded times in Eq.~\eqref{eq:rettimes} or to differences in the representation of satellite attitude.

The final accuracy of a global sphere reconstruction depends on the statistical properties of the model accuracy, in the sense that possible inaccuracies affecting some observations because of its specific geometric arrangement are likely to get ``blended'' and ``smoothed out'' by other, more accurate observations featuring a more benign configuration. It is therefore appropriate to give also a statistical assessment of these differences.

\begin{figure}[hbt]
\begin{center}
	\includegraphics[width=0.48\textwidth]{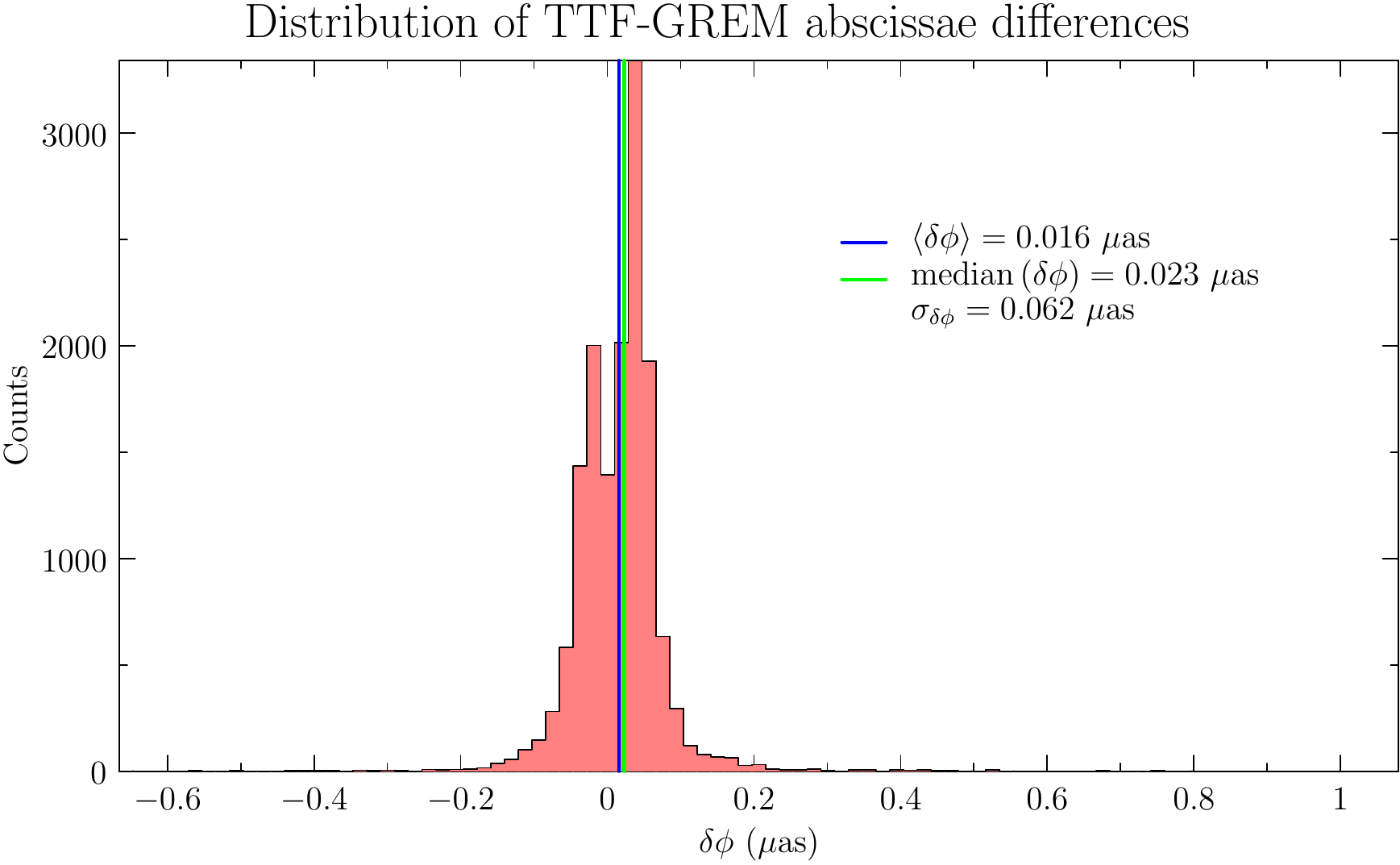}
\end{center}
\caption{Histogram of the abscissae differences between the GSR-TTF and GREM implementations. The vast majority of the observations show differences in the interval of $\pm 0.2~\mu as$, well below the accuracy required by Gaia.}
\label{fig:histoTTF}
\end{figure}

The histogram of Fig.~\ref{fig:histoTTF} shows the distribution of abscissae differences for observations taken over a period of 40 days. The vast majority of the differences are in the range $\pm 0.2 \mu as$.

%\begin{figure}[hbt]
%\begin{center}
%	\includegraphics[width=0.5\textwidth]{GSR/histo_centGSR1+.png}\hfill
%	\includegraphics[width=0.5\textwidth]{GSR/histo_codeGSR1+.png}
%\end{center}
%\caption{Histogram of the abscissae difference between the PPN-RAMOD and GREM models. The splitting emphasize the number of observations within $\pm 10 \mu as$ and with differences of more than $10 \mu as$. The chart being reasonably symmetric, only the negative values side is shown.}
%\label{fig:histoGSR1}
%\end{figure}

% \subsection{Impact of the quadrupole terms} \lb{sec:J2terms}
% %%%%%%%%%%%%%%%%%%%%%%%%%%%%%%%
% One of the expected outcome of Gaia is an improved measurement of the quadrupole moment of Jupiter~\citep{2006CQGra..23.4853C}.
% We test our implementation for the quadrupole terms by computing the deflection at a set of locations close to Jupiter and comparing the results with what expected by~\citet{2003AJ....125.1580K}.

% \red{...}

% In this sense, we show that there would be no theoretical limitation to include the quadrupole light deflection term in an astrometric model based on the TTF formalism.

\subsection{Application to the reconstruction of the celestial sphere} \label{sect:sphere}
%%%%%%%%%%%%%%%%%%%%%%%%%%%%%%%%%%
The main goal of the Gaia mission is to improve the quality of the stellar catalogs for coordinates, parallaxes and annual proper motions of the observed stars. 
This is done by evaluating the differences between the high-precision angular measurements made by the satellite and their analytical modeling as functions of the astrometric stellar parameters. 
%This will be done by comparing Gaia observations to those computed in the astrometric model starting from the approximate catalog data: minimizing the difference between the measured and simulated observation will provide better estimates of the astrometric parameters. This minimization,
As described in section~\ref{sect:GSRproc}, this is essentially a minimization problem obtained by solving a large and sparse system of linearized observation equations in the least-squares sense. Indeed, the problem is largely over-determined (Gaia will provide around $700$ observations for each star, see $e.g.$~\citet{2008IAUS..248..224M} or Fig.~\ref{Fig:GaiaFreqObs}) which justifies the least-squares solution of the equation system.

\begin{figure}[hbt]
    \includegraphics[width=0.5\textwidth]{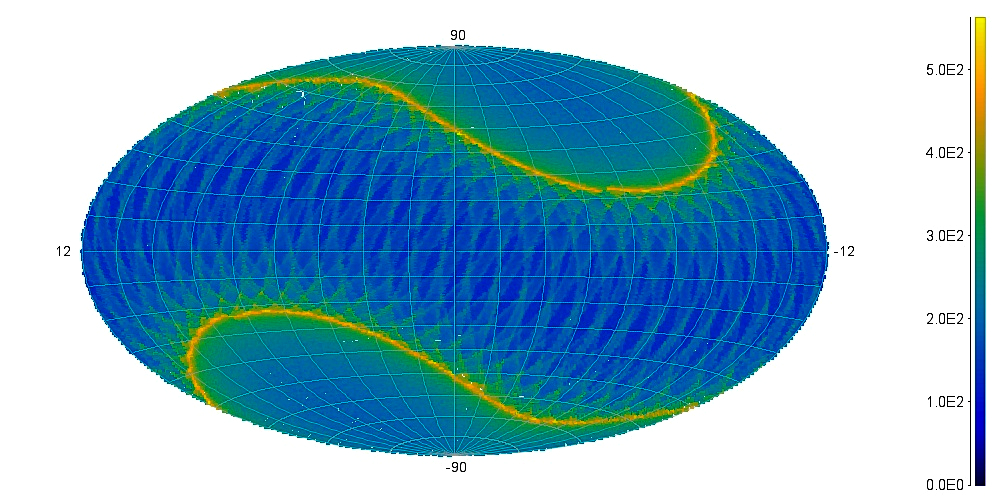}
    \caption{Frequency of observations as function of celestial coordinates due to Gaia scanning law (blue $\approx 50$ - yellow $\approx 500$)~\lb{Fig:GaiaFreqObs}}
\end{figure}

Our implementation is then tested by using observations of the Gaia satellite simulated with the GREM model, as a replacement of the real Gaia observations, and the TTF model. 
%The dataset RDS-7-F used in this test \citep{2010LL....XL-020-01} provides the needed GREM data in a form which is different from the one used for the previous comparison. 
In particular, for each observation it is possible to compute the true along-scan (AL) field angle $\eta$. Then, by definition, the abscissa of Eq.~\eqref{eq:cosphi} can be expressed as
\be
	\phi=f\frac{\Gamma}{2}+\eta,
	\lb{eq:phiGREM}
\ee
where $\Gamma$, as shown in figure~\ref{fig:astroobs_Gaia}, is the \emph{basic angle}\/, that is the angle between the axes of the preceding and following FoV, and $f=\pm1$ with sign determined by the FoV in which the star is observed. In particular $f$ takes the minus sign in the case of $\mathbf{f}_-$, and the positive one for $\mathbf{f}_+$.

The abscissae defined in this way are the values of $\phi_{\mathrm{obs}}$ in equation~\ref{eq:ktdef}, while $\phi_{\mathrm{calc}}$ is given by the TTF model. This allows the computation of the known term of the linearized observation equation. Such equation is completely set up by computing the first order
derivatives with respect to the unknown parameters, as described in section~\ref{sec:partialscompastro}, for which we still need to  define the linearization point. Finally, measurement errors must also be introduced in each observation equation.   
%. How the actual system eventually looks like, however, still requires the definition of the input values.
%with our model presented in section~\ref{sec:abscissaesimul} has two scopes: evaluating its accuracy in the specific case of the Gaia mission and setting the basis for the computation of the linearized observation equations and the reconstruction of a celestial sphere.

Specifically, the \emph{observed} abscissa $\phi_{\mathrm{obs}}$ produced from the dataset should represent the value of a \emph{measurement}. However, the simulated dataset provides us with the $\it true$ values
%what one actually gets is the \emph{true} value
, in the sense specified above, $i.e.$, the GREM model is supposed to give the correct physical representation of light propagation. Moreover, the coefficients of the linearized equation are computed at specific values, which are our best estimation of the unknowns. Once again, the input catalog provides the \emph{true} values, while in a more realistic situation one should take into account that our initial guesses will be perturbed by some catalog error. The worse these errors, the larger are the linearization errors which will eventually affect the solution.

% The observation Eq.~\eqref{eq:eqphi_lin}, contains at its l.h.s. the difference between the measured and simulated observations and at its r.h.s. the sensitivity of the observable with respect to each parameter we want to estimate. 

% However, we need to test our models and procedures before dealing with actual data.
%(1) by now we have no observational data from Gaia (it will be launched in late 2013), (2) even after launch, we will not know the "real" star parameters but only the approximation contained in the actual catalogs. 
%We should then find a way to test our models and procedures.

The above discussion, therefore, makes it clear why in our tests we follow a procedure in three steps:
\begin{enumerate}
	\item we use the \emph{true} catalog values to simulate both Gaia ``measurements'' ($\phi_{\mathrm{obs}}$) and compute the right-hand-side of Eq.~\eqref{eq:eqphi_lin}. This is equivalent to assume no measurement or catalog error on our a-priori information and allows us to test the reliability of our model (see column $0101$ in Tab.~\ref{tab:resLSQR});
	\item we apply a random gaussian error ($\mu = 0$, $\sigma_\mathrm{o}=125~\mu$as) to $\phi_{\mathrm{obs}}$, but we use the true values for $\phi_{\mathrm{calc}}$ and for the coefficients. This simulates the impact of the measurement error on the quality of the updated catalog (see column $0102$ in Tab.~\ref{tab:resLSQR}), but minimizes the linearization errors on the final solution;
	\item in addition to the above measurement error, we apply a gaussian noise of $\sigma=20$~mas to the catalog values. This introduces a contribution to the final solution due to the linearization errors of Eq.~\eqref{eq:eqphi_lin} (see column $0301$ in Tab.~\ref{tab:resLSQR}).
\end{enumerate} 
We emphasize that these kind of runs are only testing the applicability of the TTF model in the case of the estimation of the astrometric parameters. The actual case of a Gaia-like mission, the attitude parameters have to be estimated as well within the global sphere reconstruction. This is because neither the usual instrumentation for the on-board attitude reconstitution nor a short-term astrometric determination can provide a sufficiently good reconstruction, given the measurement accuracy. In this proof-of-principle, however, the main goal is to test the applicability of TTF; therefore, as a first step, it is required that we compare its accuracy with that of the Gaia baseline model GREM treating the attitude as perfectly known.
%, and to compare its accuracy with that of the Gaia baseline model, this is not only acceptable, but to a certain extent \emph{required}, as a first step.
% to produce the "real" observations $\cos \phi_{obs}$, we will first add some noise to the catalog data (in order to simulate the measuring error), while the simulated observations $\cos \phi_{calc}$ will be produced after adding a constant error to the catalog. If our procedure is well conceived, it should be able to retrieve the initial catalog value.

We therefore apply this procedure to a subset of RDS-7-F, namely, to $132\,000$ stars homogeneously distributed on the sky for which the 5 years of the planned mission provide $\sim~92,400,000$ individual observations. 
%For this scope, we use the dataset RDS7-F generated by the Gaia Coordination Unit~2 (CU2) responsible for data simulation.
%, in order to produce
%\begin{enumerate}
%	\item the measured abscissa $\cos \phi_{obs}$ with a white Gaussian noise of $\sigma=2\; mas$ previously added to the astrometric coordinates $\alpha_*$ and $\delta_*$;  
%	\item the computed abscissa $\cos \phi_{calc}$ by correcting the catalog values of a constant value of $\Delta \alpha_* = 100\; mas$ and $\Delta \delta_* = 50\; mas$.    
%\end{enumerate} 

The final goal of the procedure is to retrieve as much as possible the catalog values for the estimated parameters of the $132\,000$ processed stars. We evaluate the quality of the estimation simply by computing the residuals of the updated minus true astrometric parameters (separately for each of the five) and afterwards by calculating their average and $\sigma$.

If the two models gave exactly the same value of the abscissae, and allowing an infinite machine precision, in the $0101$ case we would expect exactly a zero solution. Since computers have a finite computing precision, and numerical errors tend to cumulate, even if the first condition would be met one should expect deviations from this solution larger than the machine precision, which in our case is about $10^{-16}$. Such numbers have to be interpreted as radians, which implies that deviations cannot be smaller than $10^{-4}~\mu$as. Moreover, the different numerical predictions of the two models, if present, can show up both as random errors and systematic biases in the TTF reconstruction, namely in non-negligible $\sigma$ and averages respectively. 
%In the Gaia sense, 
Within Gaia's astrometric tolerances, the two models are equivalent if in such test one obtains sub-$\mu$as values for these two quantities.

In the $0102$ case we introduce a purely Gaussian measurement error, which should decrease with $\sqrt{n_\mathrm{obs}}$, where $n_\mathrm{obs}$ is the number of observations of a particular star. 
%When the attitude is supposed known exactly, in fact, 
In fact, when the attitude is assumed perfectly known, along with the instrumental/global parameters, each star is completely independent from all the others, and the complete system is indeed equivalent to a set of independent systems, one for each star.
Figure~\ref{Fig:ResVSNobs} confirms the expected behavior of the residuals for each parameter and each star.
Considering that Gaia is going to measure each star about 700 times, on average, and that there are five astrometric parameters, one can roughly estimate that
\be
	\sigma_\mathrm{f}\simeq\sqrt{\frac{5}{700}}\sigma_\mathrm{o}\simeq 10~\mu\mathrm{as}.
\ee

\begin{figure}[hbt]
\begin{center}
  \begin{subfigure}[b]{0.5\linewidth}
    \centering
    \includegraphics[width=0.9\linewidth]{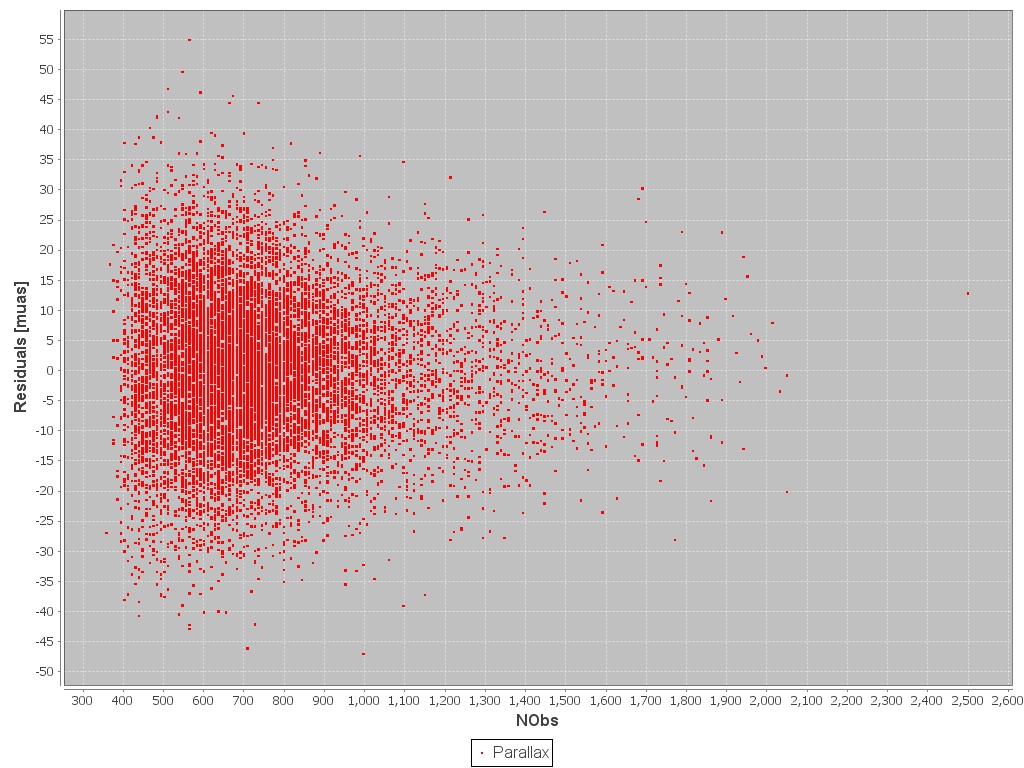} 
    \caption{$\varpi$} 
    \label{fig7:a} 
  \end{subfigure} 
  \begin{subfigure}[b]{0.5\linewidth}
    \centering
    \includegraphics[width=0.9\linewidth]{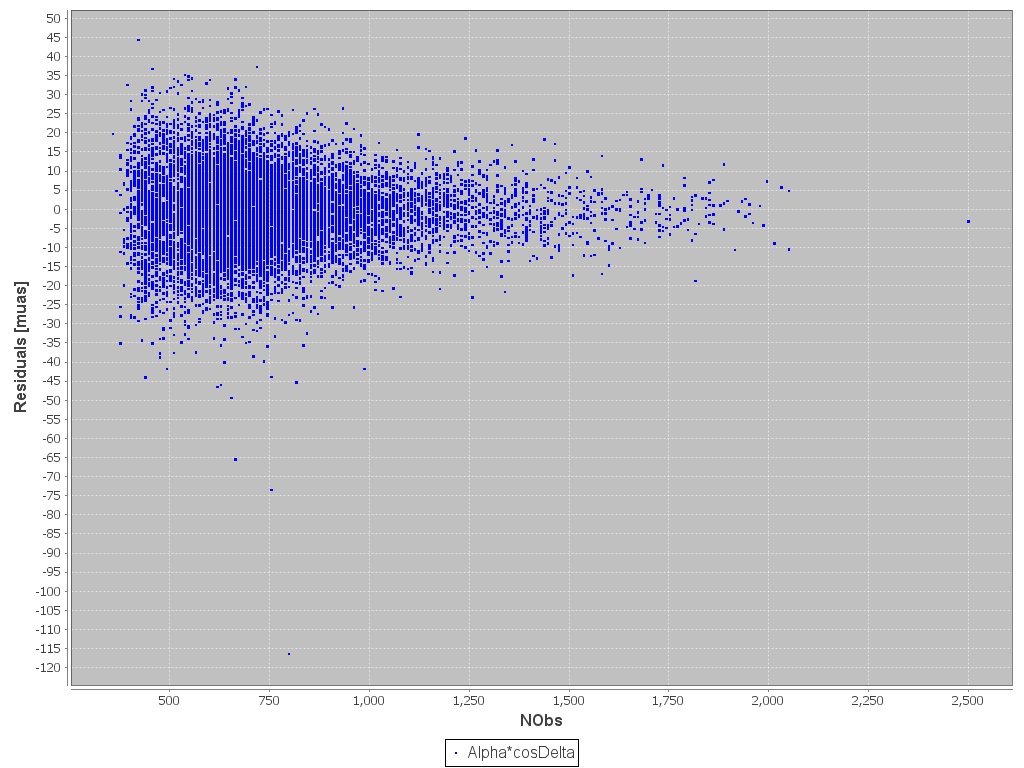} 
    \caption{$\alpha \cos{\delta}$}
    \label{fig7:b} 
    \vspace{4ex}
  \end{subfigure}%% 
  \begin{subfigure}[b]{0.5\linewidth}
    \centering
    \includegraphics[width=0.9\linewidth]{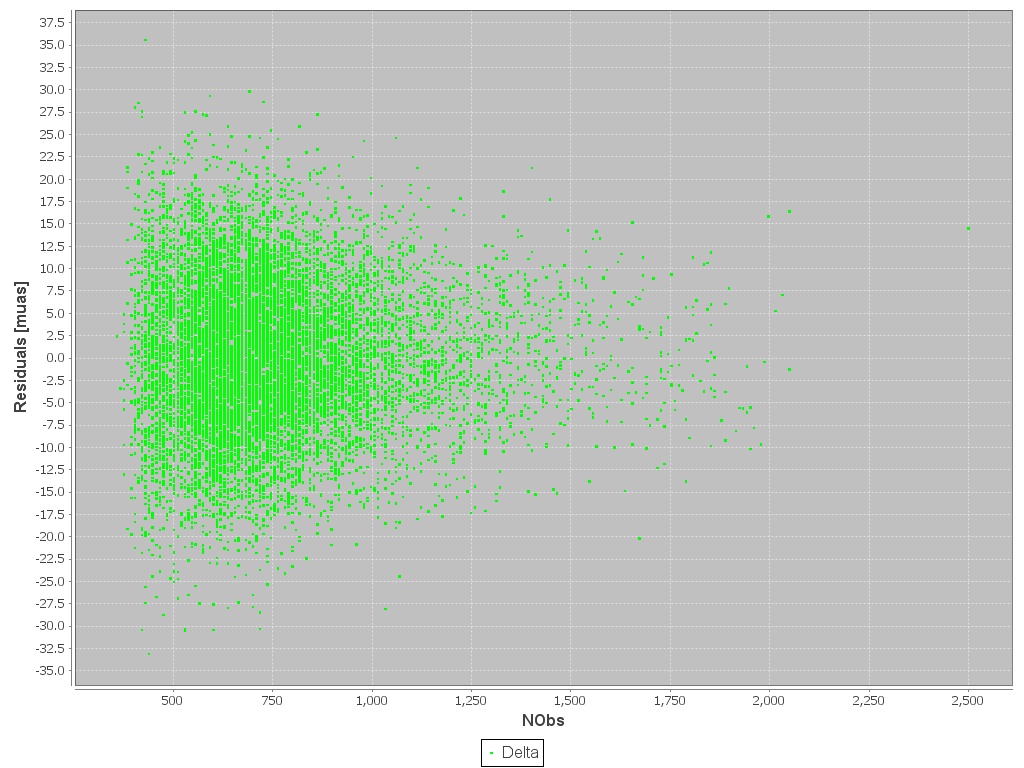} 
    \caption{$\delta$} 
    \label{fig7:c} 
    \vspace{4ex}
  \end{subfigure} 
  \begin{subfigure}[b]{0.5\linewidth}
    \centering
    \includegraphics[width=0.9\linewidth]{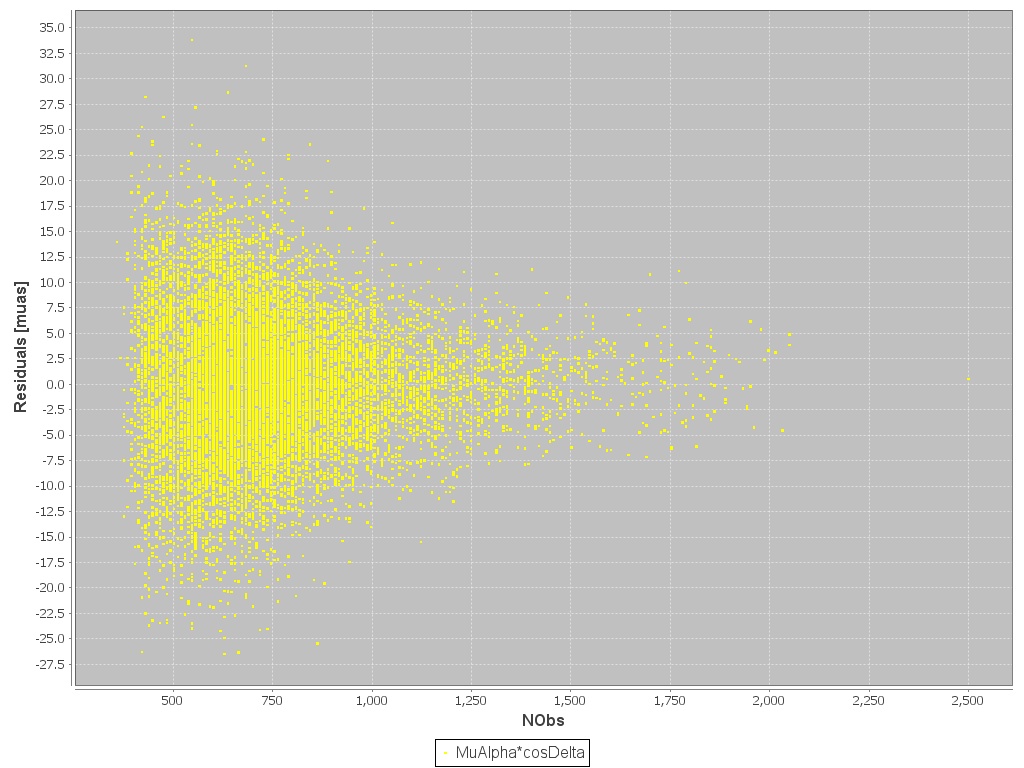} 
    \caption{$\mu_\alpha \cos{\delta}$} 
    \label{fig7:d} 
  \end{subfigure}%%
  \begin{subfigure}[b]{0.5\linewidth}
    \centering
    \includegraphics[width=0.9\linewidth]{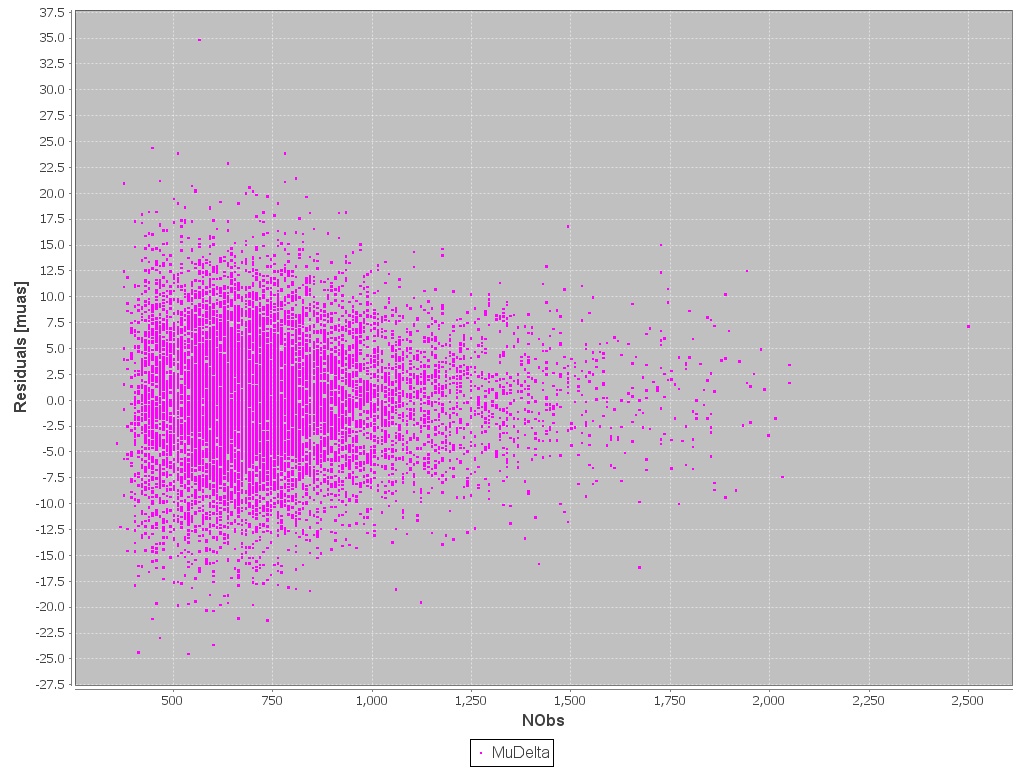} 
    \caption{$\mu_\delta$} 
    \label{fig7:e} 
  \end{subfigure} 
\end{center}
  \caption{Astrometric post-fit residuals for a set of $132,000$ stars as function of the number of observations for each star (red: $\varpi$,  blue: $\alpha \cos{\delta}$, green: $\delta$, yellow: $\mu_\alpha \cos{\delta}$, purple: $\mu_\delta$) \lb{Fig:ResVSNobs}). 
%Initial values: measure error $\sigma= 2\; mas$, catalog error $\alpha= 100\; mas$ - $\delta= 50\; mas$. After one iteration the residuals have improved of a factor $\approx 10$, depending on the number of observations for a given star.  
\lb{fig:resLSQR}}
\end{figure}

% \begin{figure}[hbt]
%     \includegraphics[width=0.5\textwidth]{Series5_Elements610990301_S_TTF_5p_ResVSobs_Cut3sigma.jpg}
%     %\includegraphics[width=0.7\textwidth]{fig7b}
%     \caption{Astrometric post-fit residuals for a set of 12000 stars as function of the number of observations for each star (red: $\varpi$,  blue: $\alpha \cos{\delta}$, green: $\delta$, yellow: $\mu_\alpha \cos{\delta}$, purple: $\mu_\delta$) \lb{Fig:ResVSNobs}}
% \end{figure}

Finally, in the $0301$ case, we have to consider the effect of the linearization errors. It is well known that, as long as the initial guess provided by the catalog is sufficiently close to the true values, this test should give the same result of the $0102$ case. It is implicit that the expression ``the same result'' has to be intended in the numerical sense, that is ``the same with respect to the required accuracy'', which in the Gaia case means to the sub-$\mu$as level.

\begin{table}
\noindent \begin{centering}
\def\arraystretch{1.2}
\setlength{\tabcolsep}{3pt}
\begin{tabular}{c|cc|cc|cc}
%\cline{2-7} 
\multicolumn{1}{c}{} & \multicolumn{2}{c|}{0101} & \multicolumn{2}{c|}{0102} & \multicolumn{2}{c}{0301}\tabularnewline
\cline{2-7} 
\multicolumn{1}{c}{} & $\bar{x}$ & $\sigma_{x}$ & $\bar{x}$ & $\sigma_{x}$ & $\bar{x}$ & $\sigma_{x}$\tabularnewline
\hline 
$\varpi$ 					& 0.000 & 0.040 & -0.059 & 11.520 & -0.033 & 11.480 \tabularnewline
$\alpha \cos{\delta}$ 		& -0.002 & 0.090 & 0.003 & 9.180 & -0.64 & 9.420 \tabularnewline
$\delta$ 					& -0.009 & 0.100 & -0.001 & 7.960 & 0.000 & 8.000 \tabularnewline
$\mu_{\alpha} \cos{\delta}$ & 0.001 & 0.020 & -0.011 & 6.440 & 0.021 & 6.430 \tabularnewline
$\mu_{\delta}$ 				& -0.001 & 0.020 & -0.014 & 5.680 & 0.010 & 5.660 \tabularnewline
%\hline 
\end{tabular}
\par\end{centering}
\center \caption{Results with GSR-TTF applied to 5 years of observations of 132K stars homogeneously distributed on the sky (in $\mu\mathrm{as}$ and $\mu\mathrm{as}/\mathrm{yr}$). \label{tab:resLSQR}}
\end{table}

\begin{table}
\noindent \begin{centering}
\def\arraystretch{1.2}
\setlength{\tabcolsep}{3pt}
\begin{tabular}{c|cc|cc|cc}
%\cline{2-7} 
\multicolumn{1}{c}{} & \multicolumn{2}{c|}{0101} & \multicolumn{2}{c|}{0102} & \multicolumn{2}{c}{0301}\tabularnewline
\cline{2-7} 
\multicolumn{1}{c}{} & $\bar{x}$ & $\sigma_{x}$ & $\bar{x}$ & $\sigma_{x}$ & $\bar{x}$ & $\sigma_{x}$\tabularnewline
\hline 
$\varpi$ 					& -0.123 & 0.270 & -0.130 & 12.02 & -0.032 & 12.06 \tabularnewline
$\alpha \cos{\delta}$ 		& -0.011 & 0.240 & -0.012 & 9.610 & 0.519 & 16.950 \tabularnewline
$\delta$ 					& -0.050 & 0.310 & -0.053 & 8.270 & -2.943 & 12.600 \tabularnewline
$\mu_{\alpha} \cos{\delta}$ & -0.001 & 0.170 & 0.024 & 6.710 & 0.753 & 18.260 \tabularnewline
$\mu_{\delta}$ 				& -0.006 & 0.150 & 0.006 & 5.860 & 1.120 & 10.690 \tabularnewline
%\hline 
\end{tabular}
\par\end{centering}
\center \caption{Results with RAMOD@GSR2 applied to 5 years of observations of 132K stars homogeneously distributed on the sky (in $\mu\mathrm{as}$ and $\mu\mathrm{as}/\mathrm{yr}$). \label{tab:resLSQRGSR1+}}
\end{table}

Table~\ref{tab:resLSQR} shows the results of these tests, while Fig.~\ref{fig:resLSQR} shows the post-fit residuals distribution for all astrometric parameters.
% \red{(Add $\chi^2$ estimation (and skewness?) to test the gaussianity of the distribution?)}
%According to the first one, therefore, for what concerns the astrometric unknowns 
Therefore, for what concerns the astrometric unknowns, we can conclude that the GREM and TTF models are equivalent. Moreover, TTF is able to recover these parameters down to the expected level of accuracy. Finally, starting from a reasonable catalog error the same astrometric accuracy can be recovered at the sub-$\mu$as level.

Table~\ref{tab:resLSQRGSR1+} reports the results of analogous runs on the same simulated datasets, but this time obtained using the RAMOD@GSR2 model, the one operational in the current version of the GSR system (GSR2) at DPCT.
Runs 0101 and 0102 compare reasonably well with the corresponding ones of TTF, although the sub-$\mu$as level of the average residuals already reveals the expected  intrinsic lower accuracy of the RAMOD@GSR2 model.
The discrepancies appearing in run 0301, the one reflecting the real case scenario, are significant at the $\mu$as level and we carefully addressed their origin. 

%These data compare with those of Table~\ref{tab:resLSQRGSR1+}, containing the results of the corresponding tests made using the GSR1+ model, namely the one which is currently operational in GSR. The accuracy of the latter is worse than the $\mu$as level needed for Gaia, and this reflects in the numbers given in the table. 
The ultimate reason of this discrepancy is actually non-linearity effects
%, which become larger in GSR1+ rather than TTF, and 
which can be recovered 
% in the former with a further solution, 
by means of a so-called ``iteration for non-linearity'', as described in detail in a forthcoming paper~\citep{AVDemonRun17}.

\begin{figure}[hbt]
\begin{center}
  \begin{subfigure}[b]{0.5\linewidth}
    \centering
    \includegraphics[width=0.9\linewidth]{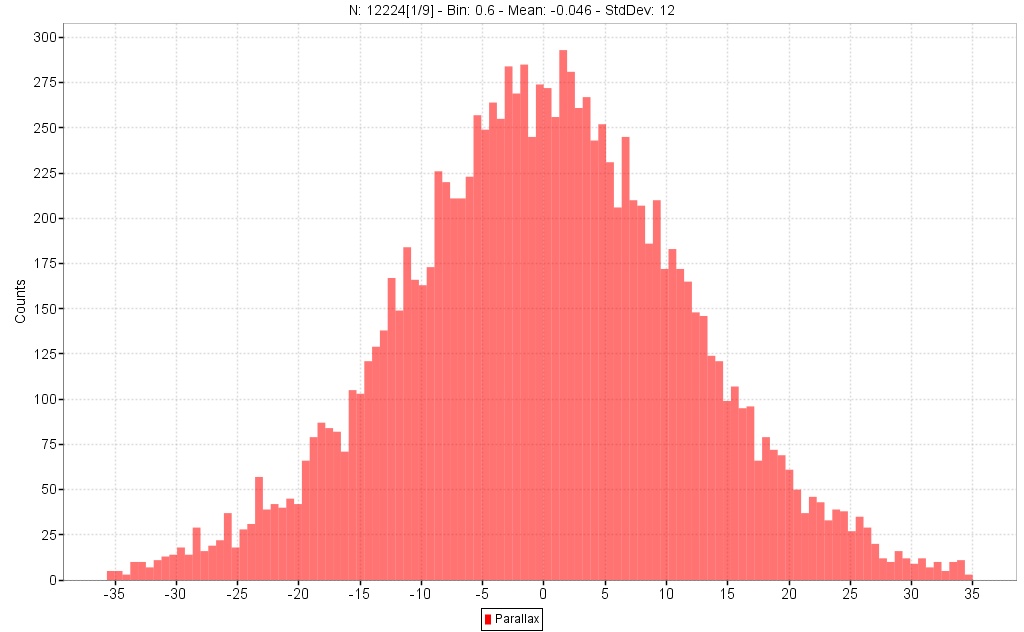} 
    \caption{$\varpi$} 
    \label{fig7:a} 
  \end{subfigure} 
  \begin{subfigure}[b]{0.5\linewidth}
    \centering
    \includegraphics[width=0.9\linewidth]{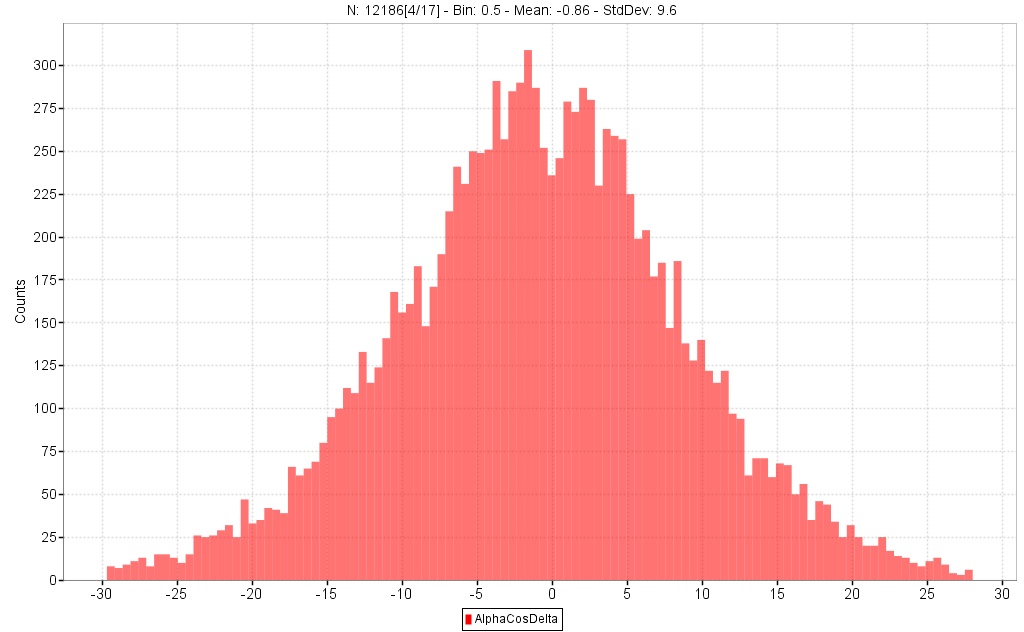} 
    \caption{$\alpha \cos{\delta}$}
    \label{fig7:b} 
    \vspace{4ex}
  \end{subfigure}%% 
  \begin{subfigure}[b]{0.5\linewidth}
    \centering
    \includegraphics[width=0.9\linewidth]{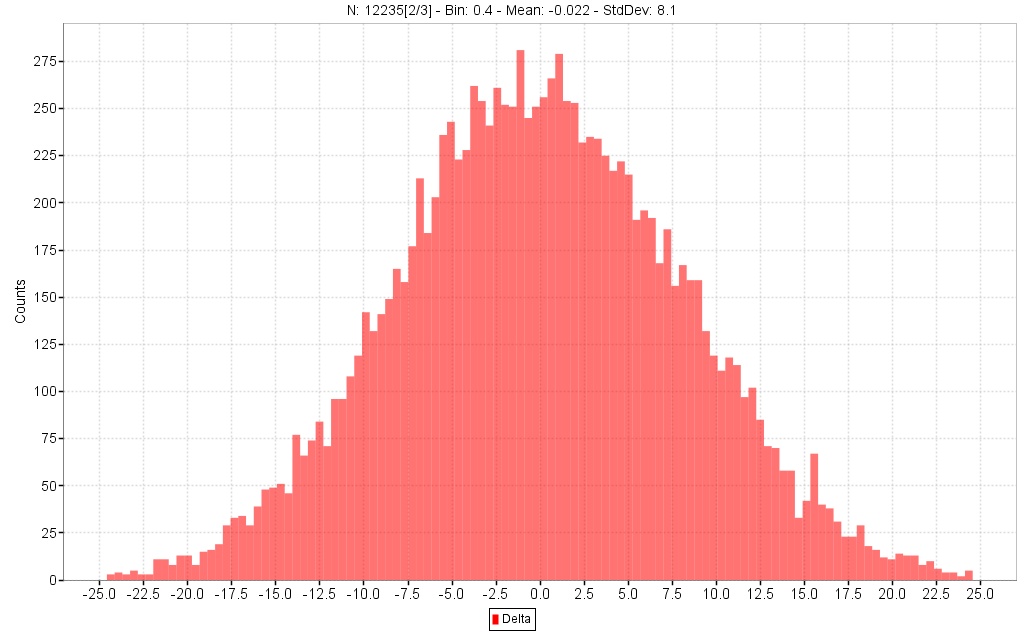} 
    \caption{$\delta$} 
    \label{fig7:c} 
    \vspace{4ex}
  \end{subfigure} 
  \begin{subfigure}[b]{0.5\linewidth}
    \centering
    \includegraphics[width=0.9\linewidth]{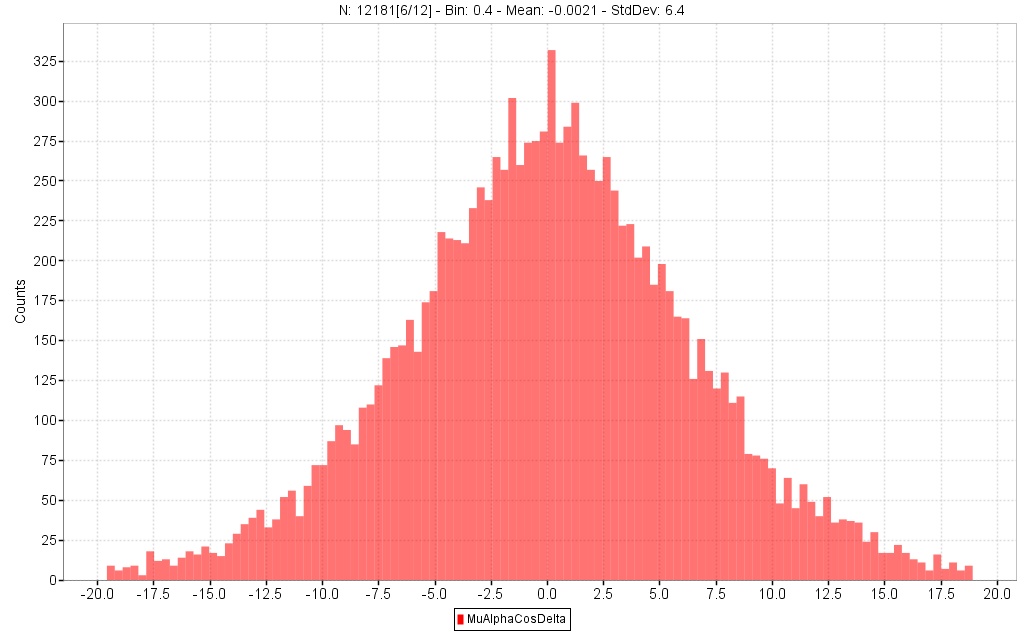} 
    \caption{$\mu_\alpha \cos{\delta}$} 
    \label{fig7:d} 
  \end{subfigure}%%
  \begin{subfigure}[b]{0.5\linewidth}
    \centering
    \includegraphics[width=0.9\linewidth]{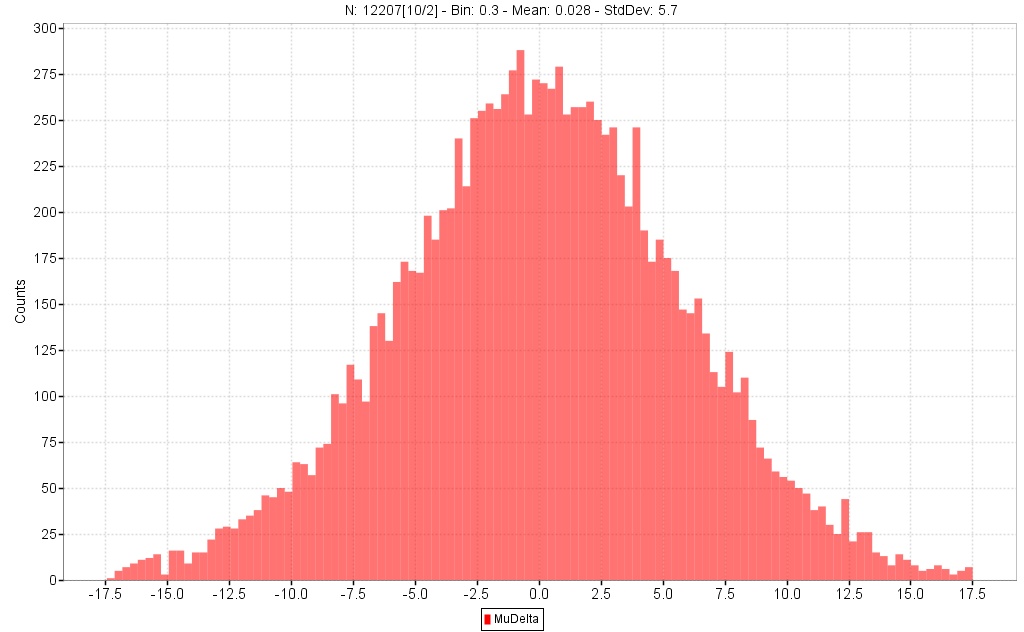} 
    \caption{$\mu_\delta$} 
    \label{fig7:e} 
  \end{subfigure} 
\end{center}
  \caption{Histogram of post-fit residuals for the astrometric parameters affected by measurement and catalog errors (see 0301 column in Table~\ref{tab:resLSQR}). 
%Initial values: measure error $\sigma= 2\; mas$, catalog error $\alpha= 100\; mas$ - $\delta= 50\; mas$. After one iteration the residuals have improved of a factor $\approx 10$, depending on the number of observations for a given star.  
\lb{fig:resLSQR}}
\end{figure}

\subsection{Sensitivity analysis for the global PPN parameter $\gamma$} \label{sect:sensgamma}
%%%%%%%%%%%%%%%%%%%%%%%%%%%%%%%%%%%
The problem in a real Gaia-like sphere reconstruction is more complex. As already mentioned above, it includes the simultaneous estimation of satellite attitude. Besides attitude, a determination of some instrument parameters is also necessary to obtain an unbiased solution. These include, for example, long-term variations of some instrument parameters, or their fine calibration to a level which cannot be reached by daily processes. The latter, however, are unrelated to the relativistic astrometric model, and are therefore out of the scope of this paper.

%Another unknown which \emph{is} related to the relativistic model, instead, is 
On the other hand, the $\gamma$ parameter of the Parametrized Post-Newtonian formulation $is$ indeed part of the relativistic model. It belongs to a class of unknowns of the sphere reconstruction problem which is called \emph{global} because it enters in every single observation equation. 
%Solving a system with the astrometric and attitude unknowns would require more powerful machines and more complex algorithms, and it involves a part of the relativistic model which is different from the one which deals with the light propagation, which is the scope of the TTF model. It is therefore out-of-scope here, and it will be presented in a forthcoming publication. On the contrary, the $\gamma$ parameter is surely connected with the problem of light propagation. Furthermore, the problem of solving the astrometric parameters together with the PPN-$\gamma$ is sufficiently simple to be treated with simple algorithms and not so powerful machines.
In this case, the least-squares estimation cannot be reduced to the solution of $N$ independent $5\times5$ systems of equations
solving for the corrections to the catalog values of the $5$ astrometric parameters of each of the $N$ stars. Since the 
normal matrix is now block diagonal with $1$-column border, it must be solved using a more complex algorithm.
It has been shown (see, $e.g.$,~\cite{2010A&A...516A..77B}) that the problem can be tackled by first solving the reduced normal equations for 
the $\delta \gamma$ unknown, and then forwarding that solution into the $N$ diagonal blocks to solve for the astrometric parameters. 
Since for each simulated sphere solution we only have one estimate of gamma, we can study its statistical properties by using a Monte-Carlo (MC) simulation. 

%In order to evaluate the performance of our model in estimating the PPN-$\gamma$ parameter from Gaia observations, we perform a MonteCarlo simulation constituted of $100$ combined solutions of both the PPN-$\gamma$ and the $5$ astrometric parameters. Each system of equations is built from the observations of $12\,000$ stars spanning the nominal $5$-years duration of the Gaia mission. 
We performed a MC experiment by running $100$ simulations for the simultaneous solution of both PPN-$\gamma$ and the $5$ astrometric parameters. For this experiment the stellar sample (and the corresponding ensemble of satellite simulated observations) extracted from RDS-7-F had to be set to $12,000$ primary sources because of limited resources allocated by the GSR2 system to the development (and test) infrastructure.    
Every MC solution provides an estimation $\gamma_\mathrm{e}$ of the PPN parameter~\footnote{Actually, each least-squares solution provides the adjustment $\delta \gamma_{e}$ to the $\gamma$ catalog value $\gamma_{cat}$ in the sense $\gamma_{e}=\gamma_{cat}+\delta \gamma_{e}$.}, which can be considered a random extraction from the Gaussian sample of $\gamma$ estimations. The statistical accuracy in the determination of this parameter is thus given by the standard deviation of the distribution of the residuals with respect to the true value $\gamma_\mathrm{t}$ ($=1$, the GR value), namely of the variate $\Delta\gamma=\gamma_\mathrm{t}-\gamma_\mathrm{e}$.

The catalog errors for all runs were set:
\begin{itemize}
	\item for the astrometric quantities, to the same values as for the 0301 test in Sec.~\ref{sect:sphere};
	\item for the PPN-$\gamma$, to 10$^{-3}$. 
\end{itemize}

Following the analysis presented in~\citet{2003A&A...399..337V}, it is possible to give an order-of-magnitude estimation of the expected accuracy on a single $\gamma$ measurement by applying error propagation to the formula of gravitational light deflection (e.g.,~\cite{gravitationBook}).
%using the approximation that the relative accuracy on $\gamma$ is of the same order of that on the deflection effect $\delta\alpha$, namely
%\[
%	\frac{\sigma_{\delta\gamma}}{\gamma}\sim\frac{\sigma_\alpha}{\delta\alpha}.
%\]

In the case of a Gaussian measurement error of $125~\mu$as, as in our simulations, and a most favorable observational configuration at $45^\circ$ from the Sun, the propagated error on $\gamma$ is  
$\sigma_\gamma \simeq 2.5 \times 10^{-2}$.
%In our case, since we apply a random Gaussian error of $\sigma_\alpha=125~\mu$as as in the previous simulations, we can roughly consider each observation as a $\sigma_{\delta\gamma}^\mathrm{o}\simeq2.5\cdot10^{-1}$ estimation of $\sigma_{\delta\gamma}/\gamma$. 
Assuming that each observation only contributes to the estimation of $\gamma$, and considering that the $12,000$ primaries  are observed more than $8~10^{6}$ times in $5$ years, the improvement factor of 
$2.8\times10^{3}$ yields a best-accuracy value of
$\sigma_\gamma \simeq 1 \times 10^{-5}$.
%Moreover, as a first approximation, one can consider each star \emph{almost} independently. Under this condition, each star provides on average about $720$ measurements, which have to be used to determine $6$ parameters (the five astrometric ones plus the $\gamma$). These figures give a number-of-observations to number-of-unknowns ratio $Q=120$, which implies that each star can in principle provide an estimation of $\gamma$ with an accuracy of $\sigma_{\delta\gamma}^*\simeq\sigma_{\delta\gamma}^\mathrm{o}/\sqrt{Q}=2.3\cdot10^{-3}$. Finally, using the observations of all the $12\,000$ stars gives a further improvement factor of about $1.1\cdot10^2$, thus yielding a final estimation of $\sigma_{\delta\gamma}\simeq2.1\cdot10^{-5}$.

%As shown in Fig.~\ref{Fig:gamma}, t
The outcome of the MC simulation 
%of $100$ runs 
is a distribution of residuals $\Delta\gamma=\gamma_\mathrm{t}-\gamma_\mathrm{e}$ centered at $-6.1\times 10^{-5}$ and with a standard deviation $\sigma_{\gamma MC} = 1.05 \times 10^{-4}$.

Despite the apparent order-of-magnitude discrepancy with the estimated accuracy above, it can be shown that this result is in fair agreement with actual expectations.
Indeed, it is well known that the $\gamma$ parameter is highly correlated with parallax, which is also estimated in our sphere reconstructions. For Gaia such correlation is similar to that of Hipparcos, namely $\rho\simeq-0.92$ \citep{2003A&A...399..337V}. This means that the above prediction for $\sigma_\gamma$ should be increased by a factor $\left(1-\rho^2\right)^{-1/2}\simeq2.6$, giving a revised value of $\sigma_{\gamma}\simeq 2.6\cdot10^{-5}$, i.e.\ a factor 3 better than $\sigma_{\gamma MC}$.

In has to be noted, however, that the previous analysis does not take into account the apportionment of the measurement error between 
$\gamma$ and the other stellar parameters.
Moreover, the straightforward error propagation on the light deflection formula is based on a best-case scenario that neglects other sources of errors like, e.g., a non-linear effect on the $\gamma$ relative error due to the angular measurement. 
For these reasons, the factor of 3 discrepancy above appears quite reasonable.

Finally, following again the findings in~\citep{2003A&A...399..337V}, we can extrapolate the previous value for 
$\sigma_{\gamma MC}$ to the case of a realistic number of observed stars. At Gaia magnitude $G=14$~\footnote {$i.e.$, the stellar magnitude corresponding to the level of observing error we simulated.} the number of primary stars is expected in the order of millions; if we then apply their improvement factor for the case of $1 \times 10^6$ 
$G=14$ stars we get 
$\sigma_{\gamma MC}=1.05 \times 10^{-4} /80 \simeq 1.3 \times 10^{-6}$, a value that compares well with more recent re-estimations of the Gaia potential for the measurement of PPN-$\gamma$~\citep{2010IAUS..261..306M,2012Ap&SS.341...31D}.
\section{Conclusions and perspectives} \label{sect:concl}
%%%%%%%%%%%%%%%%%%%%%%%%%%%%%%%%%%
In this paper we present the latest developments of GSR-TTF, an implementation of  the Time Transfer Functions relativistic model  within the current version of the Global Sphere Reconstruction software infrastructure for the reduction of Gaia astrometric observations (GSR2) running at the DPAC Data Processing Center in Torino (DPCT).
%a new processing tool for astrometric observations. The application presented here to a Gaia simulated dataset can be seen as a specific example of its application.

%This first implementation of GSR-TTF has been provided in the context of the GSR software, chosen by the Gaia DPAC to provide the validation of the final Gaia catalog. Therefore, it fully profits from years of developments and testing and from its modular structure.

Specifically, we provide the first results of GSR-TTF as applied to 5 years of DPAC simulated observations.

Light propagation modeling in GSR-TTF is based on the fully relativistic (post-Minkowskian) background of the Time Transfer Functions. This means it can be easily expanded to take into account smaller relativistic effects on light propagation if a higher accuracy is necessary. For instance, the impact of moving axisymmetric bodies could be easily implemented by adding their formulation as given, $e.g.$, in~\citet{2014PhRvD..90h4020H}.

We first provide analytical equations for both relativistic light propagation and aberration due to Gaia attitude and motion.
Then, we show that GSR-TTF results for the direction cosines, $i.e.$ the observed direction, of a set of sources are equivalent to those of the GREM implementation in GaiaTools (the model and software responsible for the processing of Gaia observations, see, $e.g.$,~\cite{2009LL....PB-015-02,2009LL....MTL-019-01}) at the $\mu as$ level or better.

The main goal of the Gaia mission is the best ever determination of positions, parallaxes and annual proper motions of all detected stars.
%in the Solar System neighborhood. 
Hence, we tested the ability of GSR-TTF to perform such a task on $\sim 132,000$ stars by using DPAC simulated observations over 5 years. The statistical analysis of the results proves GSR-TTF is able to recover astrometric parameters with the expected accuracy when dealing with both realistic measurement and catalog errors.
The estimate of instrumental parameters as well as corrections to the Gaia nominal attitude have not been included in this study but are available as part of the GSR framework and software infrastructure and will be experimented with the upcoming versions of the GSR-TTF implementation.

Next, we addressed the combined estimate of astrometric parameters with the global PPN-parameter $\gamma$ via a Monte-Carlo experiment.
Our results appear quite consistent with the most recent works on measuring PPN-$\gamma$ via Gaia's astrometry. %~\citep{2010IAUS..261..306M,2012Ap&SS.341...31D}.
%and showed satisfactory preliminary results by using a Monte-Carlo approach.

GSR-TTF is then a powerful processing tool to contribute to the Gaia final catalog, $e.g.$ as additional validation resource within the Global Sphere Reconstruction system of AVU. 
This has already been the case, having allowed to uncover the need of an iteration for non-linearity when solving the sphere with the lower accuracy RAMOD model (RAMOD@GSR2) currently in the operational infrastructure at DPCT.
%Moreover, being based on the TTF formalism allows it to adapt to future projects requiring a more accurate determination of relativistic light propagation, $e.g$ data analysis focusing on the analysis of few observations to test General Relativity in the PPN framework and beyond.

\begin{acknowledgements}
      The authors acknowledge the financial support of UIF/UFI (French-Italian University) program, CNRS/GRAM, CNES, and of the ASI contract to INAF 2014-025-R.1.2015 (Gaia Mission - The Italian Participation in DPAC).
      B.~Bucciarelli wishes to thank the support of the Chinese Academy of
Science President's International Fellowship initiative (PIFI) for 2017.
      We wish to thank our DPCT colleagues for their constant support throughout this work.
      We also thank DPAC-CU2 for making available the Gaia simulated data used in this paper.
\end{acknowledgements}

\bibliography{BiblioCNAP} % your references in file: Yourfile.bib
\bibliographystyle{aa} % style aa.bst

\end{document}